%% file: Seymour.tex
\documentclass[graybox]{svmult}

\usepackage{mathptmx}       
\usepackage{helvet}         
\usepackage{courier}        
\usepackage{type1cm}        
\usepackage{makeidx}         
\usepackage{graphicx}        
\usepackage{multicol}        
\usepackage[bottom]{footmisc}
\usepackage{subfigure,cite}
\definecolor{maroon}{RGB}{165,42,42}
\definecolor{fuchsia}{RGB}{238,0,238}
\renewcommand{\d}{\mathrm{d}}

\makeindex             

\begin{document}

\title*{Monte Carlo Event Generators} 
\author{Michael H. Seymour and Marilyn Marx}
\institute{Michael H. Seymour \at School of Physics and Astronomy, University of Manchester, Manchester M13 9PL, U.K. \\ \email{michael.seymour@manchester.ac.uk}
\and Marilyn Marx \at School of Physics and Astronomy, University of Manchester, Manchester M13 9PL, U.K. \\ \email{lynn.marx@hep.manchester.ac.uk}}
%
%
\maketitle


\abstract{Monte Carlo event generators are essential components of almost all experimental analyses and are also widely used by theorists and experiments to make predictions and preparations for future experiments. They are all too often used as ``black boxes'', without sufficient consideration of their component models or their reliability. In this set of three lectures we hope to open the box and explain the physical bases behind the models they use. We focus primarily on the general features of parton showers, hadronization and underlying event generation.}

\section{Motivation and Overview}
\input{Section1}
\label{sec:1}

\section{Parton showers}
\input{Section2}
\label{sec:2}
         
\section{Hadronization}
\input{Section3}
\label{sec:3}
        
\section{Underlying Event}
\input{Section4}
\label{sec:4}

\pagebreak
\begin{acknowledgement}
We thank the organizers of the 69th Scottish Universities' Summer School in Physics for the invitation to give these lectures, for organizing a very enjoyable and stimulating school, and for the putting competition, whisky tasting and ceilidh. The lectures have been developed from similar courses given over the years at several summer schools and MHS thanks members of the CTEQ and MCnet collaborations in particular for their feedback and encouragement.
\end{acknowledgement}


\input{referenc}

\end{document}

%% file: Section1.tex
Monte Carlo (MC) event generators are very widely used, especially by experimentalists in analyses but also by many theorists, who use them to make predictions for collider experiments and to develop techniques to propose to the experiments. MC are extremely important tools in High Energy Physics but unfortunately they are often used as ``black boxes'' whose outcome is treated as data. The aim of these lectures is to explain the physics behind event generators, which are mostly common between event generators but some differences will be highlighted. 

\begin{figure}
\centering
\subfigure[\label{fig:Hgg}]{\raisebox{10pt}{\includegraphics[width=.53\textwidth]{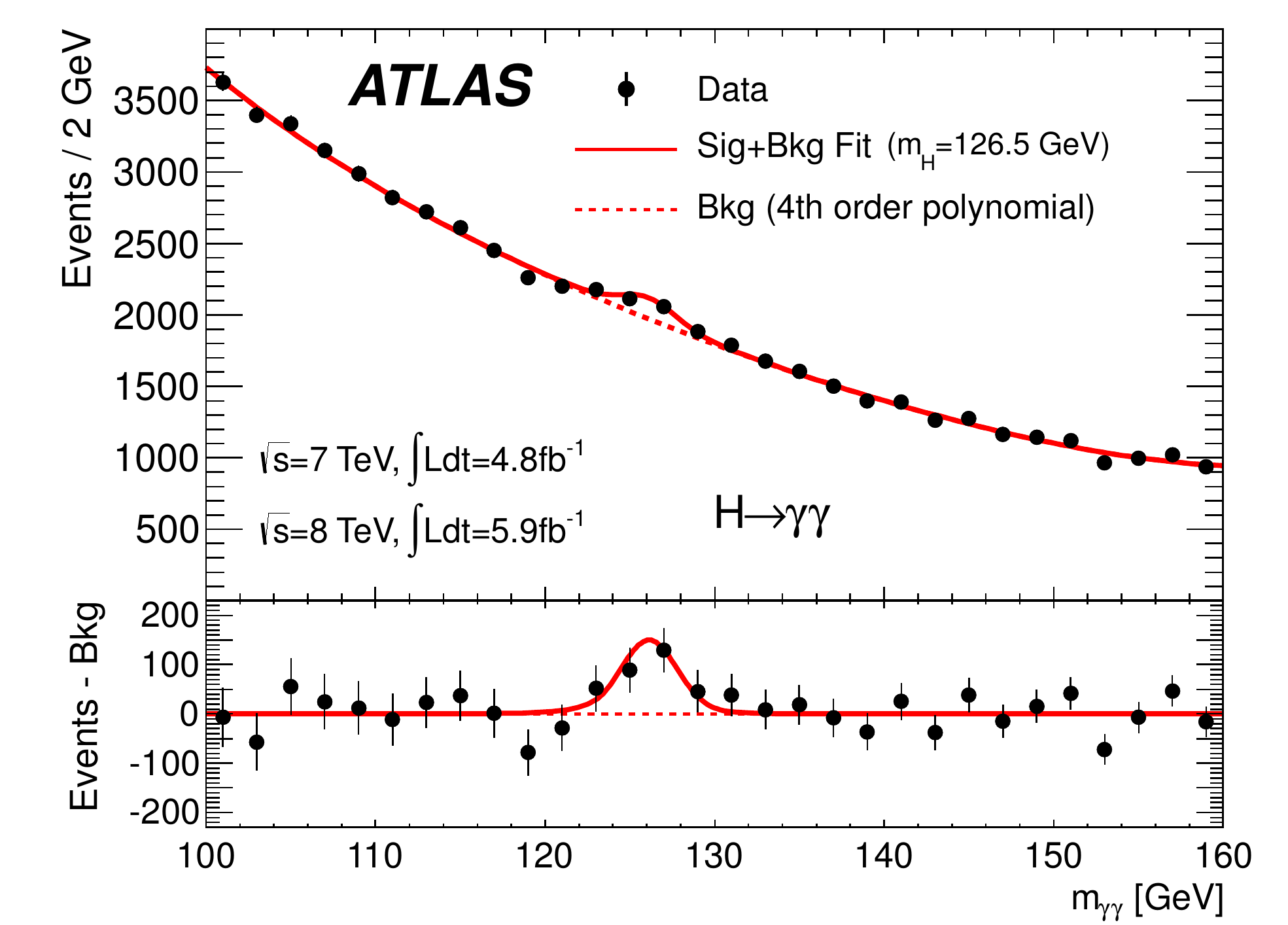}}}
\subfigure[\label{fig:Hgg2}]{{\includegraphics[width=.46\textwidth]{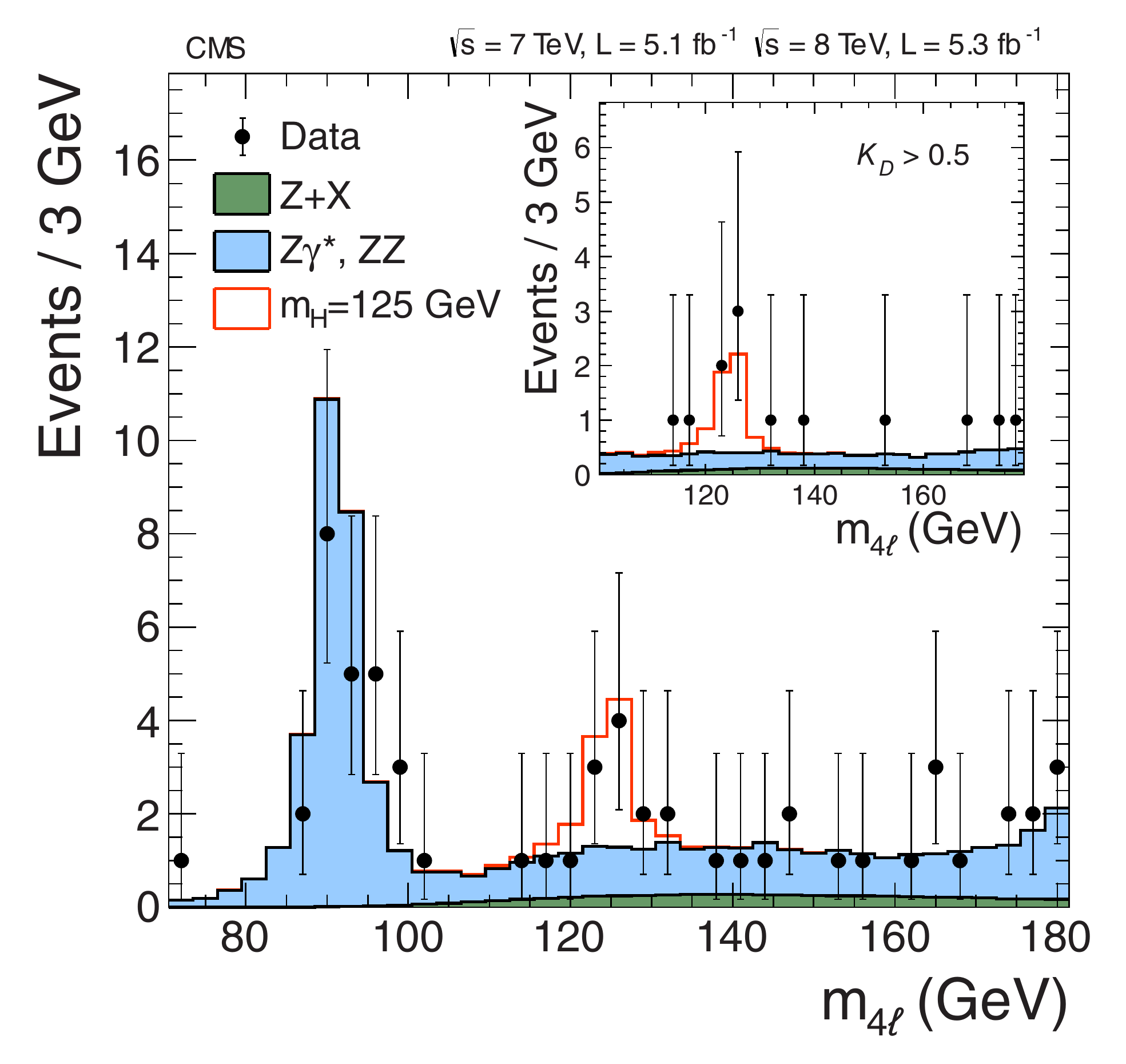}}} 
\caption{Invariant mass distributions of (a) ATLAS $H\to\gamma\gamma$ and (b) CMS $H\to ZZ\to4\ell$ candidates for the combined $\sqrt{s} = 7$ TeV and $\sqrt{s} = 8$ TeV data samples. Reproduced from \cite{Aad:2012tfa,Chatrchyan:2012ufa}.}
\end{figure}

As an example of the importance of MC, the majority of the recent Higgs discovery plots rely very strongly on MC predictions, to set limits on Higgses in certain parameter space regions as well as to discover them. This should be motivation enough to show that we need event generators for doing discovery as well as precision physics. Figure~\ref{fig:Hgg} shows the ATLAS diphoton invariant mass distribution consistent with a Standard Model Higgs boson of 126 GeV. One might ask if event generators are really still necessary when a distinct bump such as this one is visible. The answer is certainly yes, for example to quantify the significance of such a resonance and understand what particle it is. In the $H\to\gamma\gamma$ channel the resonance sits on a very steeply falling background where event generator predictions might be less important but all other discovery channels rely extremely heavily on event generators. Figure~\ref{fig:Hgg2} shows the CMS four lepton invariant mass distribution from the ``Golden Channel'', where MC predictions are crucial for signal and background modelling. 

\begin{figure}
\centering
\includegraphics[width=.65\textwidth,height=0.6\textwidth]{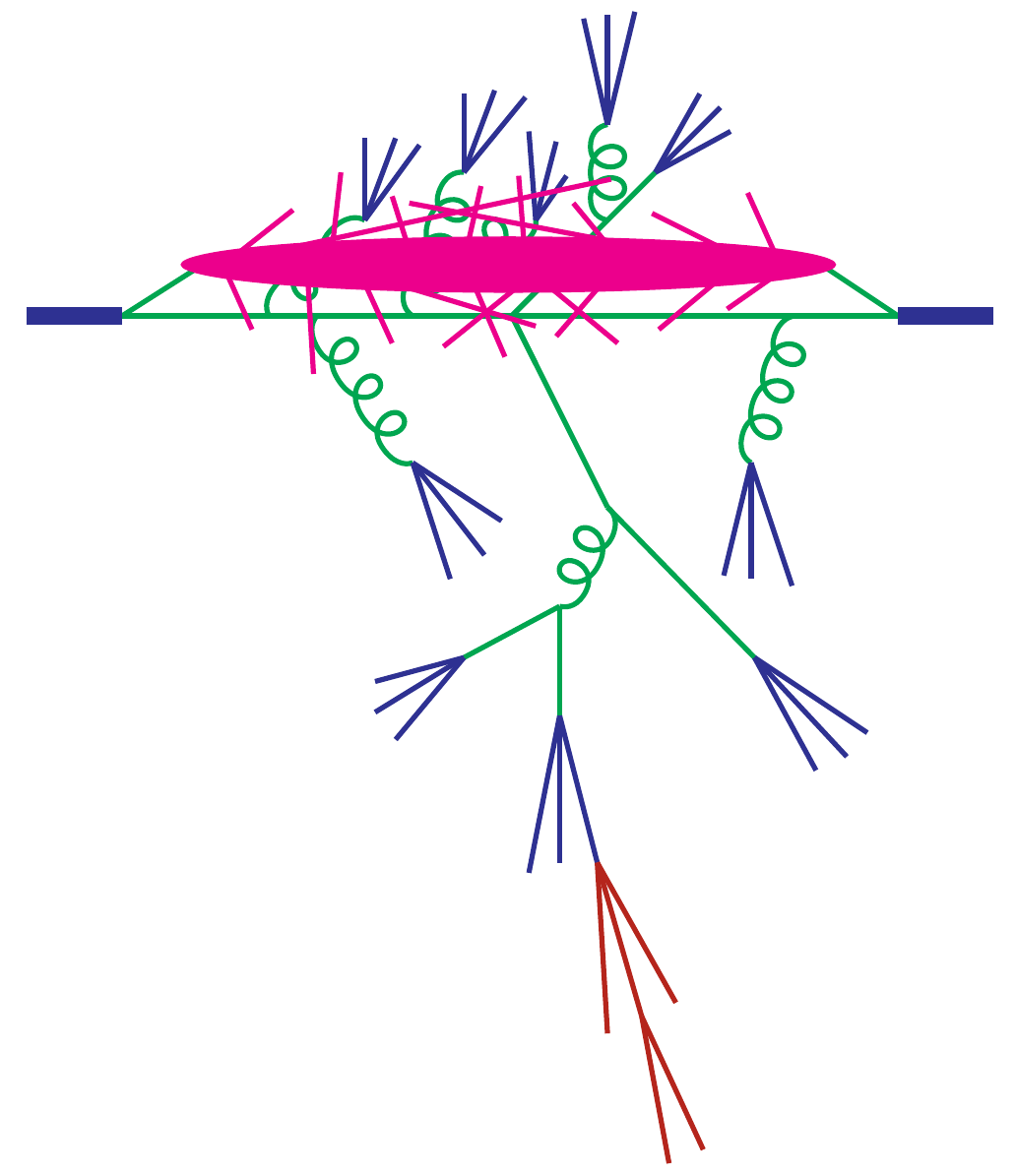}
\caption{Diagram showing the structure of a proton-proton collision, where the different colours indicate the different stages involved in event generation.}
\label{fig:event}
\end{figure}

The structure of a proton-proton collision at the Large Hadron Collider (LHC) as built up by event generators can be described by a few main steps.
These are illustrated in Figure~\ref{fig:event} where two protons come in from either side and make a collision. The colour coding corresponds to the steps into which most event generators divide the process:
\begin{enumerate}
\item Hard process
\item \textcolor{green}{Parton shower}
\item \textcolor{blue}{Hadronization}
\item \textcolor{fuchsia}{Underlying event}
\item \textcolor{maroon}{Unstable particle decays}
\end{enumerate}
The first thing an experimentalist notices when studying proton-proton collisions is that most of them are ``boring'' in the sense that only a few soft hadrons are produced and most of the event goes out along the beam pipe direction. Only a tiny fraction of events contain a high momentum-transfer process of interest. It is therefore not feasible to simulate all possible proton-proton collisions but the simulation needs to be structured with a focus on deciding what hard process is wanted (a bit like triggers at experiments which decide which events to write to tape and which to discard). 

This is done by starting the simulation at the heart of the collision and calculating from perturbation theory the probability distribution of a particular \emph{hard scatter}, which is the highest momentum transfer process in the event. Simulating the hard process is relatively straightforward because Parton Distribution Functions (PDFs) describe partons coming into the process and lowest order perturbation theory gives a probabilistic distribution of the outgoing partons. 

A more interesting stage of event generation comes from asking what happens to the incoming and outgoing partons involved in the hard collision. This is described by the \emph{parton shower} phase of event generators. The partons involved in the hard process are coloured particles, quarks and gluons. From Quantum Electrodynamics (QED) it is well known that scattered electric charges radiate photons, this is what is called Bremsstrahlung. In the same way, scattered colour charges radiate gluons and this happens for partons on their way in and out of a collision. The main difference to QED is that, due to the non-Abelian structure of $SU(3)$, gluons themselves are coloured and so an emitted gluon can itself trigger new radiation. This leads to an extended shower and the phase space fills up with (mostly) soft gluons. The parton shower can be simulated as a sequential step-by-step process that is formulated as an evolution in momentum transfer scale. The parton shower evolution starts from the hard process and works downwards to lower and lower momentum scales to a point where perturbation theory breaks down. 

Here it is necessary to switch to \emph{hadronization} models, which take account of the confinement of a system of partons into hadrons, which are seen in the detector. As well as the confinement of the produced partons, it is important to remember that the initial, uncoloured proton has had a coloured parton taken out if it and so it has been left in a coloured state. To get an idea of the space time structure of a collision, consider the fact that in a proton's own rest frame it is a spherical bound state, but in the lab frame the two protons are moving towards each other at very high speed and the Lorentz contraction flattens them into extremely thin pancakes. The collision happens at a point where these flat discs are completely overlapping each other in space time and so there is a very high probability that there will be other interactions apart from the hard interaction. This gives rise to the \emph{underlying event}, which is made up of secondary interactions between proton remnants. It produces soft hadrons everywhere in the event, which overlie and contaminate the hard process that was already simulated. 

The last component of event generation, which is usually not discussed in as much detail, is the fact that many of these hadrons are not stable particles but heavy resonances that then go on to decay. A lot of improvement has been made in the last five years to model these secondary particle decays. 

Although some details differ, this brief overview of a process from hard collision to stable hadrons is effectively used by all current general purpose event generators, i.e.\ \textsf{Herwig}, \textsf{Pythia} and \textsf{Sherpa}. The lectures are organized into three main parts: parton shower, hadronization as well as underlying event and soft inclusive physics models. More details can be obtained from \cite{guide}. The classic textbook on the subject is \cite{Ellis:1991qj}.

%% file: Section2.tex
The basic idea of the parton shower is to set up in a probabilistic way a simulation of the cascade of partons that is produced by the colour charges that are accelerated in a scattering process, or created or annihilated in a pair creation process. The simulation of final state radiation (FSR) will be discussed, namely what happens to the partons as they leave the hard collision. Finally, it will be shown that the main ideas for FSR can also be applied to initial state radiation (ISR).

\subsection{Divergence of QCD Emission Matrix Elements}

\begin{figure}
\centering
\includegraphics[width=.85\textwidth]{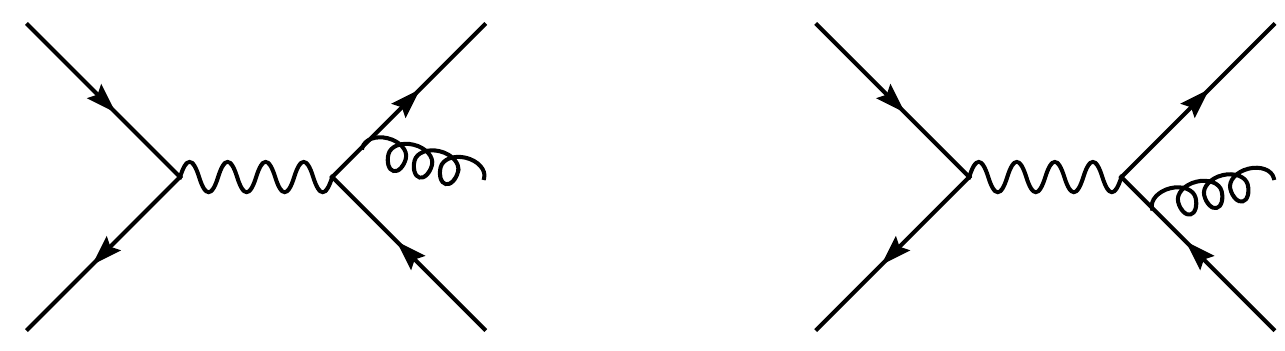}
\caption{Feynman diagrams for the process $e^+e^-\to q\bar{q}g$.}
\label{fig:Feyn}
\end{figure}

First we want to look at the simplest, non-trivial Quantum Chromodynamics (QCD) process one can study, $e^+e^-$ annihilation to jets. The tree-level cross section for $e^+e^-$ annihilation to two partons ($q\bar{q}$) is finite and, from a QCD point of view, does not have much interest. However, the first correction to this process, namely $e^+e^-$ to three partons is already very interesting. It is a good example of the more general statement that almost all QCD matrix elements diverge in different regions of phase space. It is the need to understand these divergences that will lead to the parton shower description of FSR. If we want to calculate the distribution of three partons in the final state, we need to sum two Feynman diagrams, shown in Figure~\ref{fig:Feyn}, at the amplitude level and then square them.

One can calculate the differential cross section and write it, as shown in Equation~\ref{eq:ee3p}, in terms of the opening angle $\theta$ between the quark and the gluon, the energy fraction of the gluon $z_g=E_g/E_{g,max}$, the total $e^+e^-\to q\bar{q}$ cross section $\sigma_0$, the quark charge squared $C_F$ and the QCD running coupling constant $\alpha_s\sim0.1$,
\begin{equation}
\frac{\d\sigma}{\d\cos\theta \,\d z_{g}}\sim\sigma_0\, C_F\, \frac{\alpha_s}{2\pi}\,\frac{2}{\sin^2\theta}\,\frac{1+(1-z_g)^2}{z_g}\,.
\label{eq:ee3p}
\end{equation}
This formula has several interesting features. It has a factorized form as it is proportional to $\sigma_0$ and so one can think of this as a two step process: first the $e^+e^-$ makes a two parton system, which in turn produces an extra gluon. Another important point is that it is not possible to calculate the whole probability as it is divergent in the collinear limit ($\theta\to0,\pi$)\footnote{Assuming massless quarks; the massive case will be discussed later.} and in the soft limit ($z_g\to0$)\footnote{It should be noted that here we parameterized the kinematics in the rest frame of the virtual photon, i.e.\ the rest frame of the total hadronic, three parton system. This might make it look like it is not Lorentz invariant, but one can show that the energy and angle dependence will always conspire in such a way that the final distributions are frame independent.}.
    
First we think about the physics of the collinear limit $\theta\to0,\pi$ of QCD matrix elements. Parts of the previous equation can be separated into two pieces
\begin{eqnarray}
\frac{2\,\d\cos\theta}{\sin^2\theta} &=& \frac{\d\cos\theta}{1-\cos\theta}+\frac{\d\cos\theta}{1+\cos\theta}\\
&=&  \frac{\d\cos\theta}{1-\cos\theta}+\frac{\d\cos\bar{\theta}}{1-\cos\bar{\theta}}\\
&\approx& \frac{\d\theta^2}{\theta^2}+ \frac{\d\bar{\theta}^2}{\bar{\theta}^2}\,,
\end{eqnarray}
\vspace*{-3mm}\par\noindent
where $\bar{\theta}=\pi-\theta$ is the angle between the antiquark and the gluon\footnote{This is an approximation that becomes exact in the limits of collinear or soft emission that we are interested in.}. From the middle line, we can see that we have separated this into two independent terms, the first (second) term is only divergent in the $\theta\to0$ ($\bar\theta\to0$) quark (antiquark) limit. These terms can be approximated again to expose more clearly that this is a logarithmically divergent distribution. We have written this in a manner that appears sequential: in a first step the $q\bar{q}$ pair is produced and in a second step the gluon is radiated. The probability distribution of this gluon was separated out into the sum  of two pieces, where one is associated to the quark direction and one to the antiquark direction. Rewriting the differential cross section, we can think of this as a system where each jet is evolving independently, each of which has a collinear factor:    
\begin{equation}
\d\sigma=\sigma_0\sum_{jets} C_F\, \frac{\alpha_s}{2\pi}\,\frac{\d\theta^2}{\theta^2}\,\d z\,\frac{1+(1-z)^2}{z}\,.
\label{eq:diffxsec}
\end{equation}    
Here, we have set this up in terms of the opening angle $\theta$ as it is convenient, but we could build something of the exact same form that is proportional to $\theta^2$, e.g.\ the transverse momentum of the gluon relative to the $q\bar{q}$ axis,
\begin{equation}
k^2_{\perp}=z^2(1-z)^2\theta^2E^2\,,
\end{equation}
or the total invariant mass of the quark gluon system, 
\begin{equation}
q^2=z(1-z)\theta^2E^2\,,
\end{equation}
so that
\begin{equation}
\frac{\d\theta^2}{\theta^2} = \frac{\d k^2_{\perp}}{k^2_{\perp}} = \frac{\d q^2}{q^2}\,.
\end{equation}
The choice of this variable is one of the important differences between various parton shower algorithms. In the limit of $\theta\to0$, $k_\perp\to0$, $q\to0$, all these variables give the same leading approximation to the full cross section, Equation~(\ref{eq:diffxsec}), so in describing the cross section with leading accuracy they are equivalent and this is formally a free choice. However, each involves different sub-leading corrections to the leading approximation, so the choice can, and does in practice, have important consequences for the distributions produced.
   
\subsection{Collinear Limit}
   
In Equation~\ref{eq:diffxsec}, the differential cross section was written in a factorized form $\sigma_0$ times the sum over all hard partons that are involved in the process. One can show that this is a \emph{universal} feature of QCD matrix elements and not just unique to this $e^+e^-$ case.
   
This differential cross section can be written in a universal way for an arbitrary hard process
\begin{equation}
\d\sigma=\sigma_0\, \frac{\alpha_s}{2\pi}\,\frac{\d\theta^2}{\theta^2}\,\d z\,P(z,\phi)\,\frac{\d\phi}{2\pi}\,,
\label{eq:diffxsec2}
\end{equation} 
where $z$ is the energy fraction of the parton, $\phi$ is the azimuthal angle of the splitting around the axis of the parent parton and $P(z,\phi)$ is known as the Dokshitzer-Gribov-Lipatov-Altarelli-Parisi (DGLAP) splitting kernel, which depends on flavour and spin. To give a few spin-averaged (no $\phi$ dependence) examples of the latter,
\begin{equation}
P_{q\to qg}(z)=C_F\frac{1+z^2}{1-z}\,,
\end{equation} 
\begin{equation}
P_{q\to gq}(z)=C_F\frac{1+(1-z)^2}{z}\,,
\end{equation} 
\begin{equation}
P_{g\to gg}(z)=C_A\frac{z^4+1+(1-z)^4}{z(1-z)}\,,
\label{Pgtogg}
\end{equation} 
\begin{equation}
P_{g\to q\bar{q}}(z)=T_R\left(z^2+(1-z)^2\right)\,.
\end{equation} 
       
To use the collinear limit, we do not have to take the partons to be exactly collinear, it is sufficient that the opening angle is much smaller than any angles involved in the hard process. In this limit we get this universal behaviour.

However, we still cannot describe this process probabilistically because Equation~(\ref{eq:diffxsec2}) still diverges when integrated over all possible angles. To understand the physics behind this divergence, let us take a step back and think about what we mean by having a parton in the final state of our process. As we will discuss later, partons produce jets with a finite spread of hadrons. The hadronic state produced by two exactly collinear partons is identical to that produced by a single parton with their total momentum (and colour). Without yet going into the details of the hadronization process, let us assume that there is some value of momentum of one parton transverse to the axis defined by another, below which they cannot be resolved\footnote{Note that we are not talking about our experimental ability to resolve jets, but an \emph{in principle} indistinguishability of exactly collinear partons.} and only calculate the probability distribution for \emph{resolvable} partons. That is, we introduce an arbitrary parameter, $Q_0$, which describes whether two partons are resolvable from each other or not. If $k_\perp>Q_0$, we call them resolvable and use perturbation theory to describe them as two separate particles. If $k_\perp<Q_0$ we say that they are indistinguishable from a single parton with the same total momentum. 

Now we can calculate the total probability for resolvable emission, which must be finite. Since unresolvable emission is indistinguishable from non-emission, we must add together the virtual (loop) correction to the original process and the integral of the emission probability over the unresolved region. Each is divergent, but the virtual divergence is negative and exactly cancels the real divergence. So although each of them is divergent, their sum is finite and obeys unitarity:
\begin{equation}
P(resolved)+P(unresolved)=1\,.
\end{equation} 
It is important to note that this fact is derived from Quantum Field Theory (QFT) and is not just assumed. One can encode the non-emission probability by something called a Sudakov form factor (SFF), which is a key ingredient of the MC. This SFF represents the probability that a given parton does not radiate resolvable gluons and has an exponential form. The probability of emission between $q^2$ and $q^2+\d q^2$ is
\begin{equation}
\d\mathcal{P}=\frac{\alpha_s}{2\pi}\,\frac{\d q^2}{q^2}\int_{\frac{Q_0^2}{q^2}}^{1-\frac{Q_0^2}{q^2}}\d z\,P(z)\equiv\frac{\d q^2}{q^2}\,\bar{P}(q^2)\,,
\end{equation} 
and the probability of no emission between $Q^2$ and $q^2$ is defined to be $\Delta(Q^2,q^2)$. This gives the evolution equation
\begin{eqnarray}
\frac{\d\Delta(Q^2,q^2)}{\d q^2} &=&  \Delta(Q^2,q^2)\,\frac{\d\mathcal{P}}{\d q^2}\,,\\
\Rightarrow\Delta(Q^2,q^2) &=&  \exp{-\int_{q^2}^{Q^2}\frac{\d k^2}{k^2}\,\bar{P}(k^2)}\,.
\end{eqnarray} 
This has a very similar form to the well-known formula of radioactive decay, where an atom has a (constant) probability $\lambda$ per unit time to decay:
\begin{equation}
P(\textrm{no decay after time $T$})=\exp{\,-\!\!\int^{T}\d t\,\lambda}\,.
\end{equation} 
        
The Sudakov form factor, $\Delta(Q^2,Q_0^2)\equiv\Delta(Q^2)$, represents the probability of emitting no resolvable radiation at all:
\begin{equation}
\Delta_q(Q^2)\sim\exp{-C_F\,\frac{\alpha_s}{2\pi}\log^2\frac{Q^2}{Q^2_0}}\,,
\end{equation} 
which becomes very small for large $Q^2$, reflecting the fact that a quark formed at a high scale is extremely unlikely to be unaccompanied by \emph{any} gluons.
        
\subsection{Multiple emission and running coupling}

\begin{figure}
\centering
\includegraphics[width=.85\textwidth]{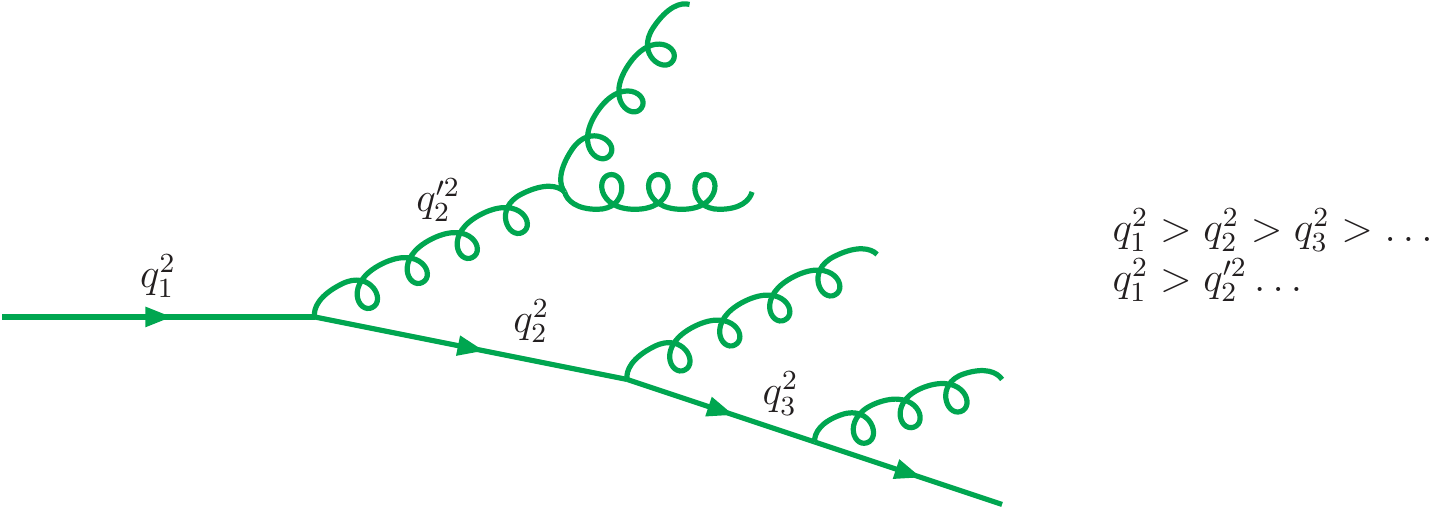}
\caption{Diagram showing multiple gluon emission off an initial quark line.}
\label{fig:multiple}
\end{figure}

We can use the universality of the DGLAP splitting function to calculate the probability that, given an initial emission, a certain gluon radiates another gluon that is more collinear than the first one. This way we can attach more and more partons, as is shown in an example in Figure~\ref{fig:multiple}. We can take the different building blocks that we have just derived, namely the tree-level splitting function and the SFF, which tells us the non-emission probability, and use them to construct the probability distribution of any complicated final state. It should be noted that one important point that needs to be specified is the initial condition. It tells us how large the initial value $q_1$ of the evolution variable can be. This is the only process dependent factor in the parton shower and we will come back to this later\footnote{One can think of this initial condition as anything that parametrizes the ``collinear-ness'' of an emission process, e.g.\ the virtuality, how far off-shell this particle was. If we do not know how a coloured particle was produced, we cannot know how it radiates.}.

To quickly touch on higher order loop corrections to emitted gluons, one can absorb a tower of higher order logarithmic corrections by replacing $\alpha_s$ by $\alpha_s(k^2_\perp)$. This is because at each higher order, $\alpha_s^n$, one encounters terms like $\beta_0^n\ln^nk^2_\perp/\mu^2$, where $\beta_0$ is the leading coefficient of the QCD beta-function and $\mu$ is the renormalization scale at which $\alpha_s$ is defined. If $\mu^2$ is very different to $k^2_\perp$ these terms are very large, spoiling the convergence of perturbation theory. But if the emission vertex is evaluated with $\alpha_s$ replaced by $\alpha_s(k^2_\perp)$, these terms are absorbed and effectively resummed to all orders. Thus, taking account of the increasing coupling at small $k_\perp$, the parton evolution is expected to fill the phase space with soft (low transverse momentum) gluons as it becomes increasingly ``easy'' (i.e.\ the probability becomes high) to emit very soft gluons. This further means that the $Q_0$ parameter is a very important physical parameter that constrains the parton shower algorithm. In order to use perturbation theory, one has to ensure that $Q_0$ is much larger than $\Lambda_{QCD}$, which is $\sim\mathcal{O}(200 \textrm{ MeV})$, and so it should be of order 1 or a few GeV.
        
\subsection{Soft limit and Angular ordering}
        
Now we want to move from the collinear limit to the soft limit, which is the other limit in which QCD matrix elements diverge. There is also a factorisation theorem for the soft limit but it has a very different form in the sense that it is universal only at the amplitude level, not at the cross section level. 

\begin{figure}
\centering
\subfigure[]{\label{fig:soft}\includegraphics[width=.45\textwidth]{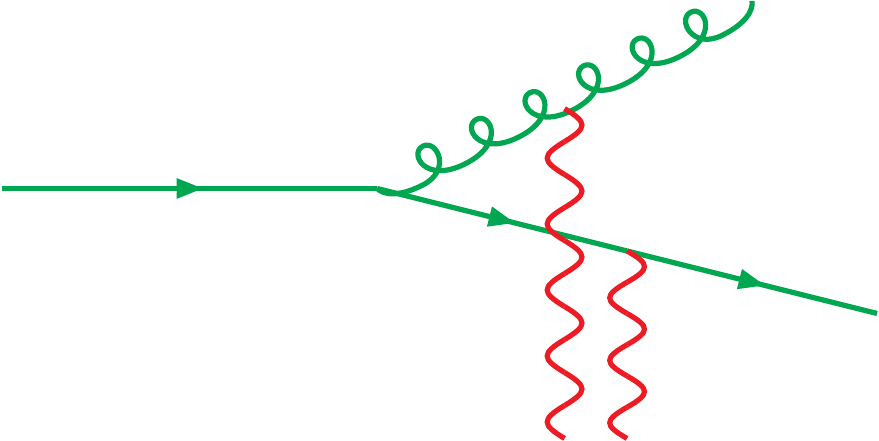}}
\hspace{\fill}
\subfigure[]{\label{fig:soft2}{\includegraphics[width=.45\textwidth]{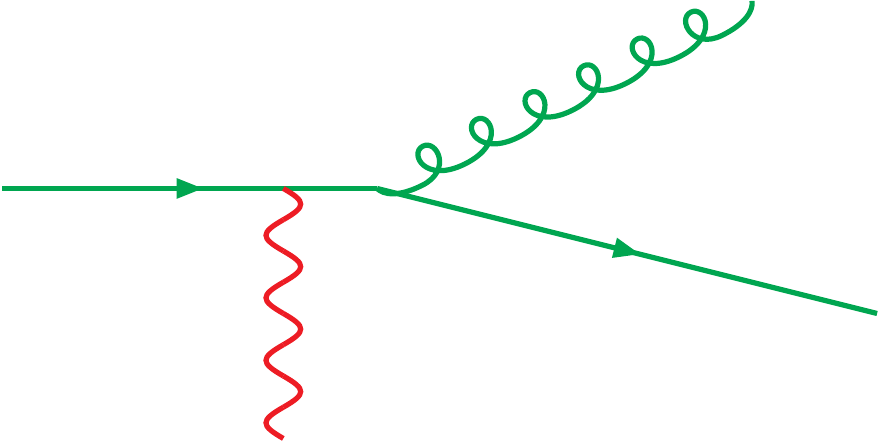}}} 
\caption{Diagram describing the soft limit: the soft gluon may be emitted by either of the outgoing partons, at the amplitude level (a), but the sum of the two diagrams is as if it was emitted by a single parton with the same total momentum and colour (b).}
\end{figure}

Consider a quark coming out of a hard process that radiates a hard gluon. We want to know what the distribution of soft gluons radiated from this system looks like. Going back to QFT, there are two possible Feynman diagrams that contribute, which we illustrate by the single diagram in Figure~\ref{fig:soft}. The soft gluon is attached to the hard gluon in one diagram and to the quark in the other. In either of these cases there is a factorisation theorem telling us that the amplitude for that process can be written as the amplitude to produce the hard gluon times a universal factor describing the radiation of the soft gluon. However, unlike in the collinear limit, we need to sum these two diagrams before we square them and since the amplitudes that they represent have similar magnitudes there will be quantum interference between them. At first this was thought to spoil the picture of independent evolution, as described before in the collinear limit. Actually this is not the case because the radiation from these two is coherent. If we sum up the two diagrams at the amplitude level and square them, the radiation pattern from this pair of partons is identical at large angles to the radiation pattern from a single quark with the same total colour charge and same total momentum as the pair of partons had if it was on-shell as shown on Figure~\ref{fig:soft2}. So at large angles, the gluons essentially only see the total colour charge, they cannot resolve the colour charges of the individual partons. On the other hand, when the opening angle to one of the hard partons is small, the corresponding single diagram dominates. We can incorporate this into our collinear parton shower algorithm by ordering in the opening angle, which will therefore correctly describe the soft limit. 

We can conclude that it is possible to construct a parton shower that can describe correctly both the collinear and the soft limits of QCD matrix elements by using the opening angle as the evolution variable and describing a wide angle gluon as if it was emitted before the internal structure of the jet has built up\footnote{At this point, we assume that the quarks are massless. We will see that the picture does not change radically for massive quarks, although for quarks heavier than $\Lambda_{QCD}$ the quark mass actually plays a similar role to the resolution scale that we have discussed, cutting off collinear emission.}.

\subsection{Initial State Radiation}
        
\begin{figure}
\centering
\includegraphics[width=.85\textwidth]{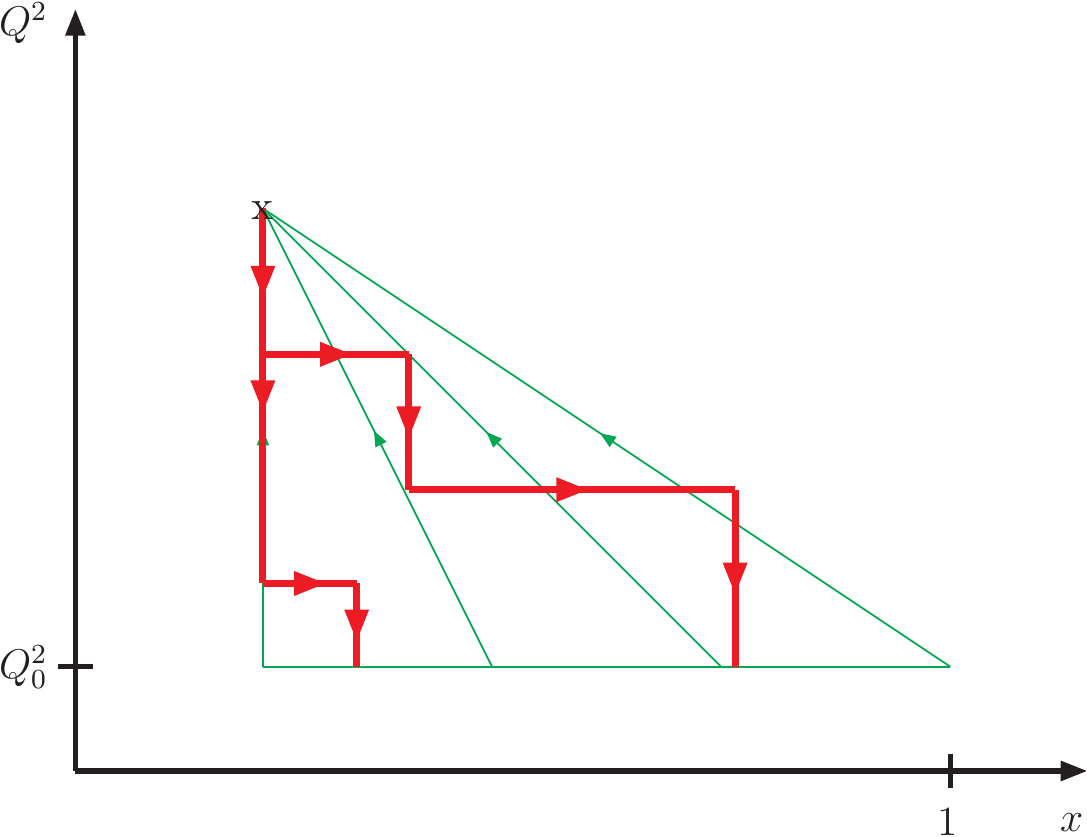}
\caption{The green lines illustrate the flow of information in analytic solutions of the DGLAP evolution equation, which yields the value of the parton distribution function at a given value of $x$ and~$Q^2$ as a function of its values at some lower value of $Q^2$, $Q_0^2$, and all higher values of $x$. The red lines illustrate typical backward evolution paths that lead to the same $x$ and $Q^2$ value: each path corresponds to one event and each corner on the path to one emitted parton.}
\label{fig:bkwdevo}
\end{figure}

FSR is fully inclusive in the sense that we want to generate the distribution of all possible parton radiation, for ISR the goal is different. Here, we want to be able to choose the hard process and ask what radiation this process is accompanied by. So even though the physics involved in ISR and FSR is essentially the same, we have certain kinematic constraints for ISR, e.g.\ we know $x$ and~$Q^2$, and we therefore do not want to generate all possible distributions but only those subject to having a fixed parton momentum at the end of the process. As illustrated in Figure~\ref{fig:bkwdevo}, we can reformulate the evolution as a \emph{backward evolution}, which probabilistically undoes the DGLAP evolution equations. We start from a particular $x$ and $Q^2$ point and work down in $q^2$ and up in $x$ towards the incoming hadron, asking progressively, what is the probability distribution of radiation that accompanies a parton of this flavour and kinematics. In the end, one finds that this algorithm is identical to FSR, but with $\Delta_i(Q^2,q^2)$ replaced by $\Delta_i(Q^2,q^2)/f_i(x,q^2)$.

\subsection{Hard Scattering and Colour Coherence}
    
We need to set the initial conditions for parton showers and here the colour coherence that we already talked about when studying the soft limit of QCD matrix elements, is important too. We take the example of quark-antiquark pair annihilation, say $u\bar{u}\to d\bar{d}$. The rules of perturbative QCD tell us that the quarks are in the fundamental representation of $SU(N_C=3)$ and the gluons are in the adjoint representation, which has $N_C^2-1$ $(=8$ for $SU(3)$) colours. Up to corrections of order $1/N_C^2$, we can think of a gluon as carrying a fundamental colour and a fundamental anti-colour label. In that approximation, the colour structure of this $q\bar{q}$ annihilation process looks completely different to the flavour structure (the ``upness'' is annihilated and becomes ``downness'', but the ``redness'' of an incoming quark gets transferred onto the $s$-channel gluon and thence onto the outgoing quark). The effect of the colour structure is best illustrated by contrasting a $u\bar{u}\to d\bar{d}$ event in which the $d$~quark goes in the forward direction of the $u$~quark with a $u\bar{d}\to u\bar{d}$ event with the same kinematics. In $u\bar{u}\to d\bar{d}$, the colour of the quark has only been scattered through a small angle and any emission from it is confined to angles smaller than the scattering angle. In $u\bar{d}\to u\bar{d}$, the colours in the initial state annihilate each other and a new colour-anticolour pair is created. Emission from both lines fills the whole of phase space. In general, the colour line of any parton can be traced through the hard process to find the parton to which it is colour connected: its colour partner. Emission from each parton is confined to a cone stretching to its colour partner and the colour coherence limits parton radiation in certain regions of phase space. 

\begin{figure}
\centering
\includegraphics[width=0.49\textwidth]{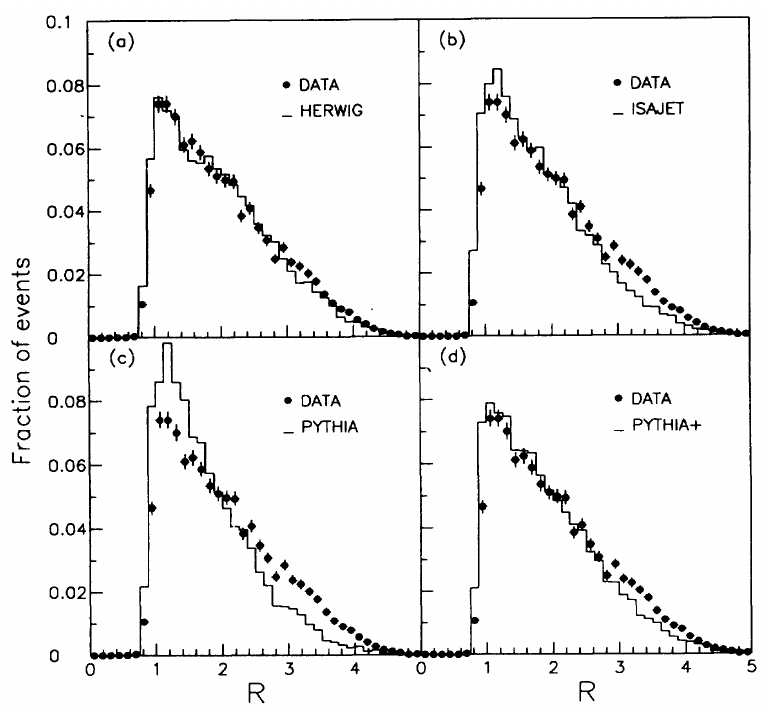}
\hspace{\fill}
\includegraphics[width=0.49\textwidth]{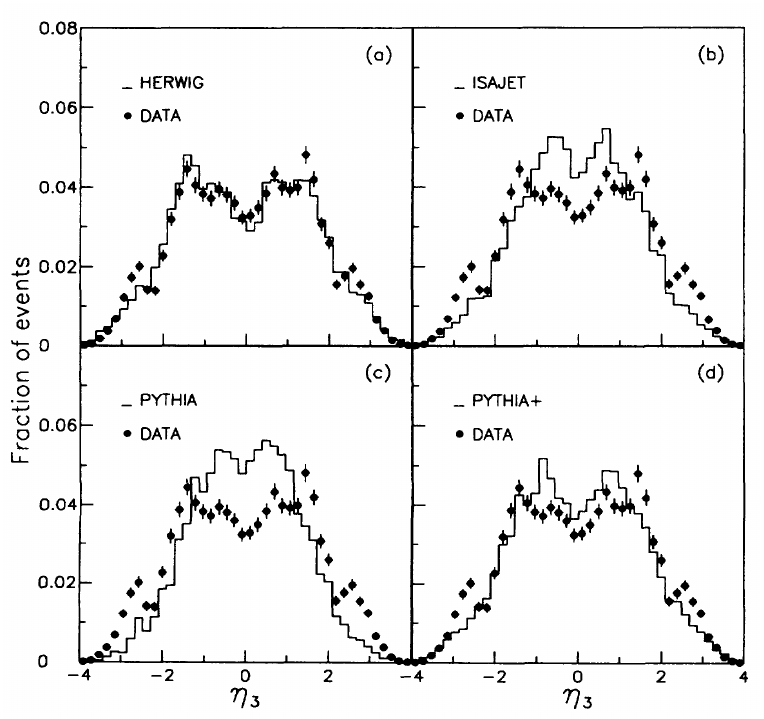}
\caption{Distributions from the CDF analysis demonstrating colour coherence in three-jet events. $R$~is the separation, in $\eta-\phi$ space, between the second and third hardest jets and $\eta_3$ is the pseudorapidity of the third hardest jet. Reproduced from \cite{Abe:1994nj}.}
\label{fig:CDF}
\end{figure}

The CDF 3-jet analysis is a classic experimental example that demonstrates that colour coherence is a real effect. In this analysis, they required two hard jets with $p_{t,1}>110$ GeV but only $p_{t,3}>10$ GeV. To map out the kinematics of this third jet, they looked at the pseudorapidity and jet-separation distributions, which can be seen in Figure~\ref{fig:CDF}. In these plots\footnote{The MC predictions here have been put through detector simulation.}, \textsf{Herwig} has colour coherence built in and predicts a dip between the hardest and second hardest jet, which was also seen by the data. The other two MCs shown, \textsf{Pythia} and \textsf{Isajet}, did not have this colour coherence built in. In this case, radiation was allowed to go everywhere and for kinematical reasons the radiation actually prefers to go in the central region. This misprediction prompted the \textsf{Pythia} authors to provide a better model, which was then called \textsf{Pythia+} and had a partial treatment of colour coherence added. This feature is nowadays part of \textsf{Pythia} by default.

\subsection{Heavy Quarks}

\begin{figure}
\centering
\includegraphics[width=0.8\textwidth]{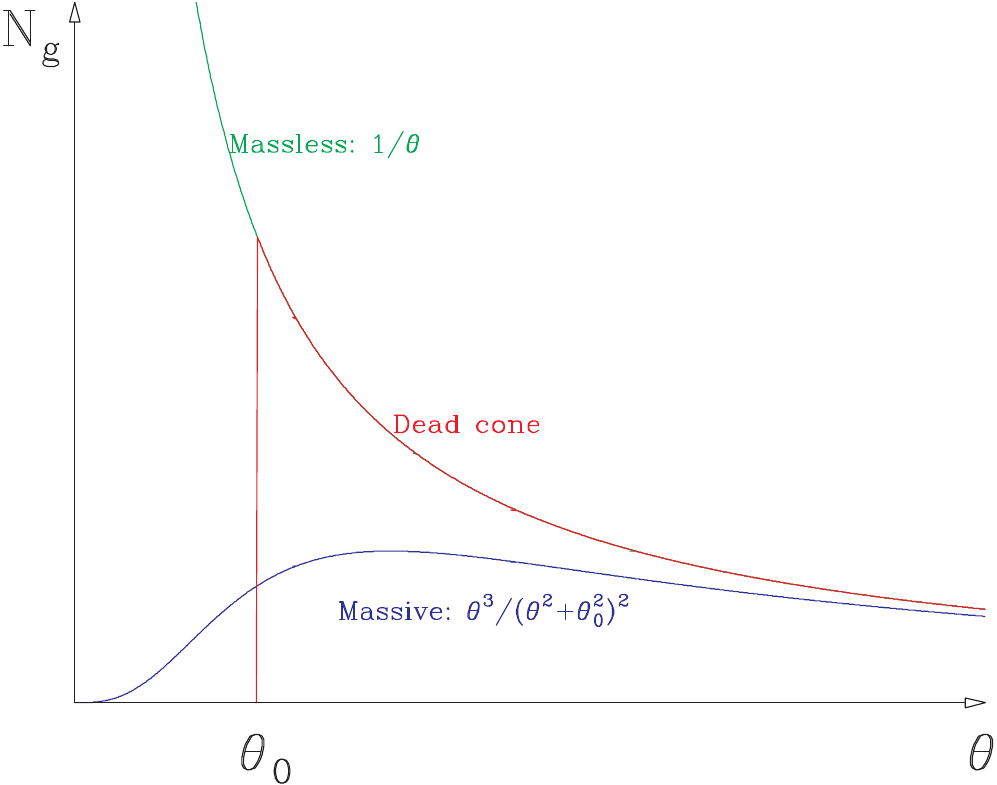}
\caption{Number of radiated gluons, $N_g$, as a function of the opening angle $\theta$ for the massless and massive quark case.}
\label{fig:heavy}
\end{figure}

So far we have only talked about the case where quarks are lighter than the confinement scale, for which their mass does not play an important role in their evolution because it is the confinement itself that provides a cut-off. But if we want to calculate the emission pattern from heavy quarks there is another colour coherence effect at work. Figure~\ref{fig:heavy} shows the emission pattern of a final state quark. The massless case, drawn in green, goes as $1/\theta$ and diverges as we go to smaller angles. The pattern from a massive quark with the same momentum, drawn in blue, becomes similar at large angle (this is again the colour coherence effect~-- wide angle gluons only see the total colour) but as we get closer to $\theta_0=\frac{m_q}{E_q}$ there is a smooth suppression and the true emission pattern turns over and goes to 0 at small angles. This means that a massive quark does not radiate at all in its forward direction. This blue curve is what is implemented in most current event generators but sometimes people still talk about the ``dead cone effect'', which was the implementation in earlier generators. It is the somewhat brutal approximation of treating the massive quark as massless up to a certain angle and then just turning off radiation in the forward direction completely. This can be shown to give the correct total amount of radiation, but can be seen to be a very crude approximation for the distribution of that radiation.

More often a quasi-collinear splitting is used\cite{Catani:2002hc,Norrbin:2000uu}, which has a smooth suppression in the forward region,
\begin{equation}
\d\mathcal{P}_{\tilde{ij}\to ij}=\frac{\alpha_s}{2\pi}\,\frac{\d\tilde{q}^2}{\tilde{q}^2}\,\d z\,P_{\tilde{ij}\to ij}(z,\tilde{q})\,.
\end{equation}
For reference, we give the splitting functions for massive quarks and spartons\footnote{Since the gluon is massless $P_{g\to gg}$ is unchanged from equation~(\ref{Pgtogg}).}:  
\begin{equation}
P_{q\to qg}(z)=\frac{C_F}{1-z}\left[1+z^2-\frac{2m_q^2}{z\tilde{q}^2}\right],
\end{equation}
\begin{equation}
P_{g\to q\bar{q}}(z)=T_R\left[1-2z(1-z)+\frac{2m_q^2}{z(1-z)\tilde{q}^2}\right],
\end{equation}
\begin{equation}
P_{\tilde{g}\to \tilde{g}g}(z)=\frac{C_A}{1-z}\left[1+z^2-\frac{2m^2_{\tilde{g}}}{z\tilde{q}^2}\right],
\end{equation}
\begin{equation}
P_{\tilde{q}\to \tilde{q}g}(z)=\frac{2C_F}{1-z}\left[z-\frac{m_{\tilde{q}}}{z\tilde{q}^2}\right].
\end{equation} 
  
\subsection{Colour Dipole Model}
    
So far we have been talking about ``conventional'' parton showers where you start from the collinear limit of QCD matrix elements and modify it to incorporate soft gluon coherence. In the colour dipole model (CDM) the starting point is somewhat different as it tries to understand the soft radiation first and to then modify that in such a way that you also get the collinear limit right. The idea is to start from the large $N_C$ approximation where a gluon is treated as a colour-anticolour pair or dipole. The emission of soft gluons from such a dipole is universal (and classical)
\begin{equation}
\d\sigma\approx\sigma_0\,C_A\,\frac{\alpha_s(k_\perp)}{2\pi}\,\frac{\d k_\perp^2}{k_\perp^2}\,\d y\,,
\end{equation}
where $y=-\log\tan\frac{\theta}{2}$ is the rapidity. In this model, we think of colour-anticolour \emph{pairs} as radiating, so the $q\bar{q}$ pair together radiate a gluon. This creates an additional colour line, separating the system into a $qg$ dipole and a $g\bar{q}$ dipole, which go on to radiate further. This way subsequent dipoles continue to cascade and instead of a $1\to2$ parton splitting, like in parton showers, you have a $1\to2$ dipole splitting, which corresponds to a $2\to3$ parton splitting.

One different feature of the CDM is that there is no explicit ISR. The hadron remnant forms a colour dipole with the scattered quark, which is treated like any other dipole, except for the fact that the remnant is an extended object. This radiates gluons, but since it is an extended object it does not radiate in its forward direction. The radiation looks like FSR from the outgoing remnant rather than ISR from the ingoing quark, but one can show that they are equivalent, with the suppression of radiation from the remnant interpreted as the suppression due to parton distribution function effects in ISR.

Most of the parton shower implementations that have appeared in the last few years~\cite{NS,DTW,Sherpa1,Sherpa2,Vincia} are based on this dipole cascade picture~\cite{CS,DK}.
    
\subsection{Matrix Element Matching}

The parton shower method is an approximation derived from QCD that is valid in the collinear and soft limits. It describes the bulk of radiation well but very often one uses event generators to search for new physics, to predict backgrounds or to model features of the signal processes and do precision physics, like e.g.\ the top mass measurement, measuring multi-jet cross sections etc. In these applications, you are not only interested in very soft and collinear emission but in systems of hard, well-separated jets. Therefore many of the applications of parton shower event generators are pushing them into regions of phase space where they are least reliable, i.e.\ away from the soft and collinear approximations and more into regions where fixed order matrix elements should describe those processes better. In order to improve their predictions, one would like to get simultaneously (at least) next-to-leading order (NLO) normalization, a good description of hard multi-jet systems but also match that with a good parton shower of the internal structure of those jets~-- i.e.\ the best of all worlds. Achieving this is known as matrix element matching and is one of the areas where MCs have developed the most in the last five years. Several methods have been proposed to combine tree-level matrix elements for several jet multiplicities simultaneously, with parton showers to describe the internal structure of the jets and the pattern of soft radiation between the jets (the ``intrajet'' and ``interjet'' event structure respectively) without double counting (the buzz-words are CKKW\cite{Catani:2001cc,Krauss:2002up} or CKKWL\cite{Lonnblad:2001iq} and MLM\cite{Alwall:2007fs} matching). Alternatively, two methods have been proposed to combine lowest-multiplicity NLO matrix elements with parton showers, again without double counting (MC@NLO\cite{Frixione:2002ik} and POWHEG\cite{Nason:2006hfa}). The current state of the art is progress towards NLO multi-jet matching\cite{NS,Lavesson:2008ah,Hamilton:2010wh,Hoche:2010kg}, which is needed for many applications at the LHC.

\subsection{Summary of available Programs}

We briefly mention some of the parton shower-related features of the different MC programs that are available.

Older programs that are still sometimes seen but not supported any more:
\begin{itemize}
\item \textsf{Pythia 6.2}\cite{Sjostrand:2000wi}: traditional $q^2$ ordering, veto of non-ordered final state emission, partial implementation of angular ordering in initial state, big range of hard processes.
\item \textsf{HERWIG}\cite{Corcella:2000bw}: complete implementation of colour coherence, NLO evolution for large x, smaller range of hard processes.
\item\textsf{Ariadne}\cite{Lonnblad:1992tz}: complete implementation of colour-dipole model, best fit to HERA data, interfaced to \textsf{Pythia} for hard processes.
\end{itemize}
Supported and new programs:
\begin{itemize}
\item \textsf{Pythia 6.3}\cite{Sjostrand:2006za}: $p_t$-ordered parton showers, interleaved with multi-parton interactions, dipole-style recoil, matrix element for first emission in many processes.
\item\textsf{Pythia 8}\cite{Sjostrand:2007gs}: new program with many of the same features as \textsf{Pythia 6.3}, many `obsolete' features removed.
\item\textsf{Sherpa}\cite{Gleisberg:2008ta}: new program built from scratch: $p_t$-ordered dipole showers, multi-jet matching scheme (CKKW) to AMAGIC++ built in.
\item\textsf{Herwig++}\cite{Bahr:2008pv}: new program with similar parton shower to \textsf{HERWIG} (angular ordered) plus quasi-collinear limit and recoil strategy based on colour flow, spin correlations.
\end{itemize}
In addition, dipole showers are available as optional plug-ins to both \textsf{Herwig++}\cite{Platzer:2011bc} and \textsf{Pythia}\cite{Vincia}.
      
\subsection{Summary}

The basic idea of parton showers is very simple: accelerated colour charges radiate gluons, but since the gluons themselves are also charged we get an extended cascade developing. This cascade is modeled as an evolution downward in momentum scale. As we approach the non-perturbative limit, we get more and more radiation and the phase space fills with soft gluons. The probabilistic language is derived from factorization theorems of the full gauge theory. Colour coherence is a fact of life: do not trust those who ignore it!

Modern parton shower models are very sophisticated implementations of perturbative QCD, but they would be useless without the hadronization model, which will be discussed next.

%% file: Section3.tex
Everything we have studied so far was based on perturbative QCD but partons are not the final state particles that come out of the collision as they cannot propagate freely. We know that hadrons are the physical final state particles, but we do not know how to calculate them, so we need a model to describe how partons are confined into hadrons~-- this is called hadronization. The two main models in use are the String Model, implemented in \textsf{Pythia}, and the Cluster Model, implemented in \textsf{Herwig} and \textsf{Sherpa}. These models will be described in more detail later but we will first look at some of the physics behind hadronization models and how they have developed.

\subsection{Phenomenological Models}

\begin{figure}
\centering
\includegraphics[width=1\textwidth]{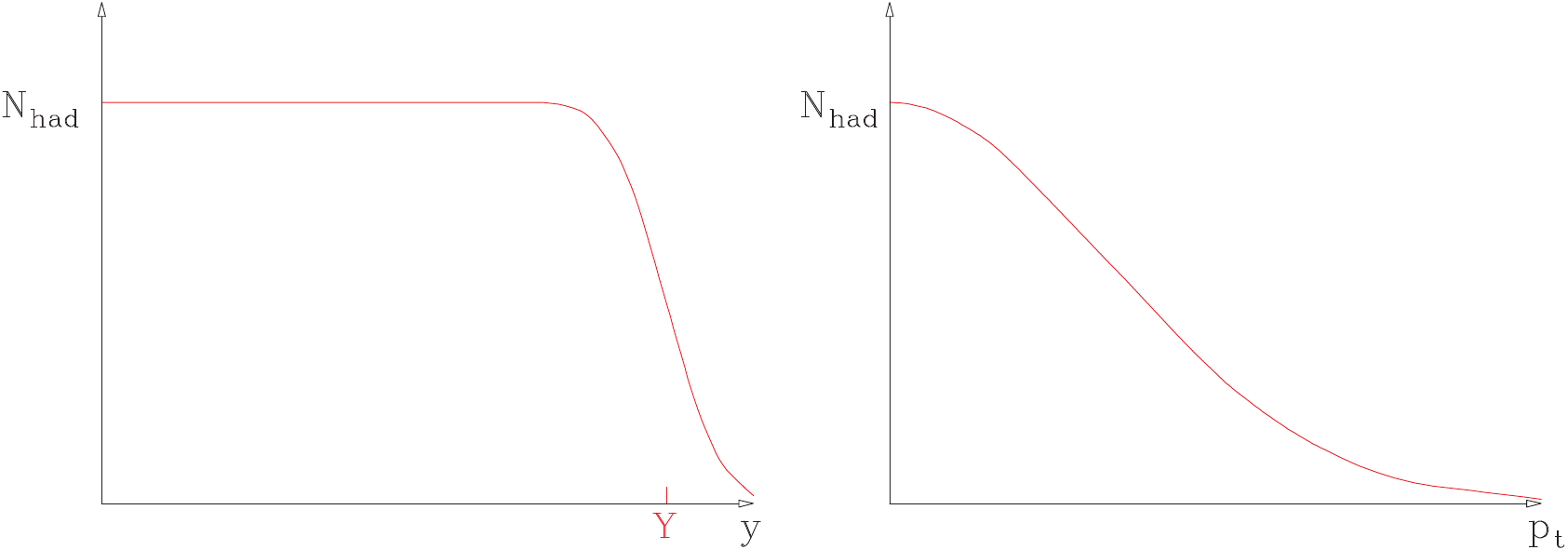}
\caption{The number of hadrons sketched as a function of rapidity, $y$, and transverse momentum, $p_t$.}
\label{fig:phenomodel}
\end{figure}

We start with an experimental observation, namely $e^+e^-$ annihilation to two jets, which is the majority of hadronic $e^+e^-$ events, and study the distribution of hadrons with respect to the axis formed by the two jets. You can measure the rapidity, $y$, and the transverse momentum, $p_t$, of hadrons relative to that axis. If you plot the number of hadrons as a function of $y$, sketched on the left in Figure~\ref{fig:phenomodel}, you find that it is roughly flat up to some maximum value and then falls off very quickly. However looking at the right side of Figure~\ref{fig:phenomodel} where the number of hadrons is sketched as a function of $p_t$, it can be seen that this distribution is roughly Gaussian with a narrow width of 1 or 2 GeV. This means that most hadrons are produced with very low $p_t$. 

One can make a very simple model based on this observation and estimate the hadronization correction to perturbative quantities. The energy of the jet is
\begin{equation}
E = \int_{0}^{Y}\d y\,\d^2p_t\,\rho(p_t^2)\cosh y = \lambda\sinh Y\,,
\end{equation}
where $Y$ is the maximum rapidity of hadrons in the jet and $\lambda$ is their mean transverse momentum, given by
\begin{equation}
\lambda = \int \d^2p_t\,\rho(p_t^2)\,p_t\,,
\end{equation} 
which can be estimated from Fermi motion where $\lambda\sim 1/R_{had}\sim m_{had}$. The longitudinal momentum can be calculated in the same way:
\begin{equation}
P = \int_{0}^{Y}\d y\,\d^2p_t\,\rho(p_t^2)\,\sinh y = \lambda(\cosh Y-1)\sim E-\lambda\,.
\end{equation} 
The jet acquires a non-perturbative mass, given by
\begin{equation}
M^2=E^2-P^2\sim2\lambda E\,,
\end{equation} 
from which it can be seen that the non-perturbative invariant jet mass is proportional to the square root of its energy. This non-perturbative component is an important contribution, e.g. a 10 GeV contribution for 100 GeV jets. Since these corrections are so important, we need a precise model to predict them.

The first set of models that were developed to describe hadronization were the so-called Independent Fragmentation Models (or ``Feynman-Field'' models) and they are a direct implementation of the procedure described above. The longitudinal momentum distribution is an arbitrary fragmentation function, a parametrization of data. The transverse momentum distribution is assumed Gaussian and the model just recursively applies $q\to q'+had$ and hooks up the remaining soft $q$ and $\bar{q}$ until the whole jet is hadronized. This model can describe $e^+e^-\to2$ jet events by construction, but has a lot of disadvantages: it is strongly frame dependent, there is no obvious relation with perturbative emission, it is not infrared safe and it is not a model for confinement.

\subsection{Confinement}

\begin{figure}
\centering
\includegraphics[width=.45\textwidth]{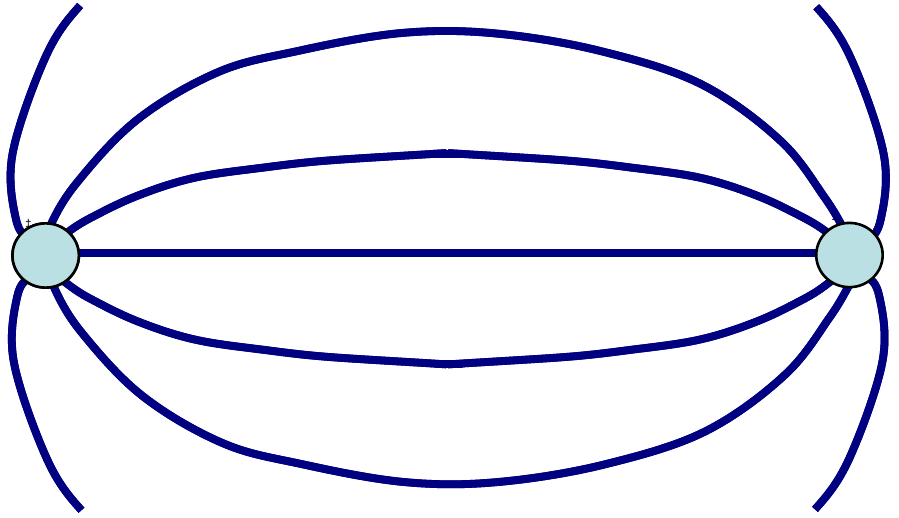}
\hspace{\fill}
\raisebox{1cm}{\includegraphics[width=.45\textwidth]{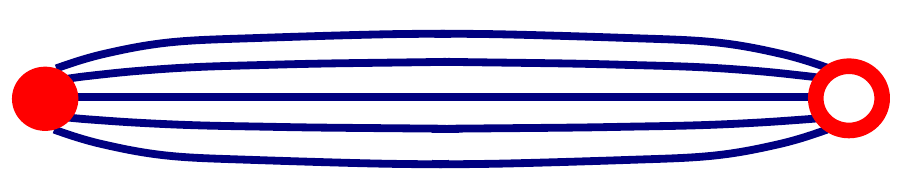}}
\caption{Field lines in QED (left) and QCD (right) between a charge and an anti-charge.}
\label{fig:confinement}
\end{figure}

We know that in QCD we have asymptotic freedom: at very short distances a $q\bar{q}$ pair becomes more and more QED-like, but at long distances the non-Abelian gluon self-interaction makes the field lines attract each other, as sketched in Figure~\ref{fig:confinement}. As two colour charges are pulled apart the field lines do not spread out and we get a constant force or a linearly rising potential. One would have to invest an infinite amount of work to pull them apart~-- this is the signal of confinement.

\begin{figure}
\centering
\rotatebox{-90}{\includegraphics[height=.8\textwidth]{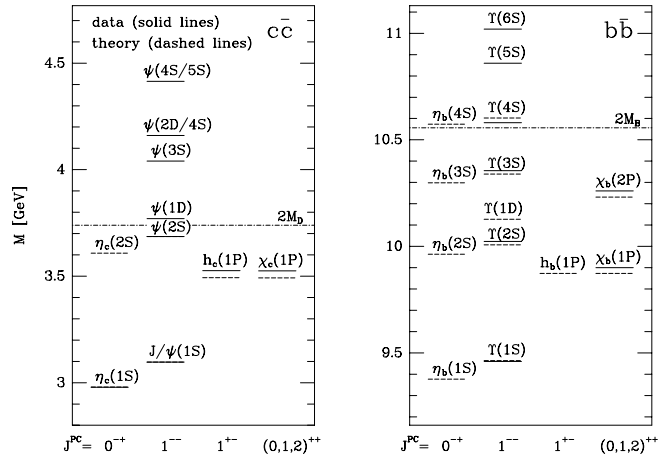}}
\vspace{-20pt}
\caption{Charmonium and bottomonium spectra, the $nS$ ($n=1,2,...$) energy levels are roughly equally spaced (in more detail, they go like $\sim n^{2/3}$). Reproduced from \cite{Ellis:1991qj}.}
\label{fig:quarkonia}
\end{figure}

This interquark potential (or string tension) can for example be measured from quarkonia spectra, as shown in Figure~\ref{fig:quarkonia}, or from lattice QCD. The string tension $\kappa$ is found to be roughly 1 GeV/fm.

\subsection{String Model}

\begin{figure}
\centering
\includegraphics[width=\textwidth]{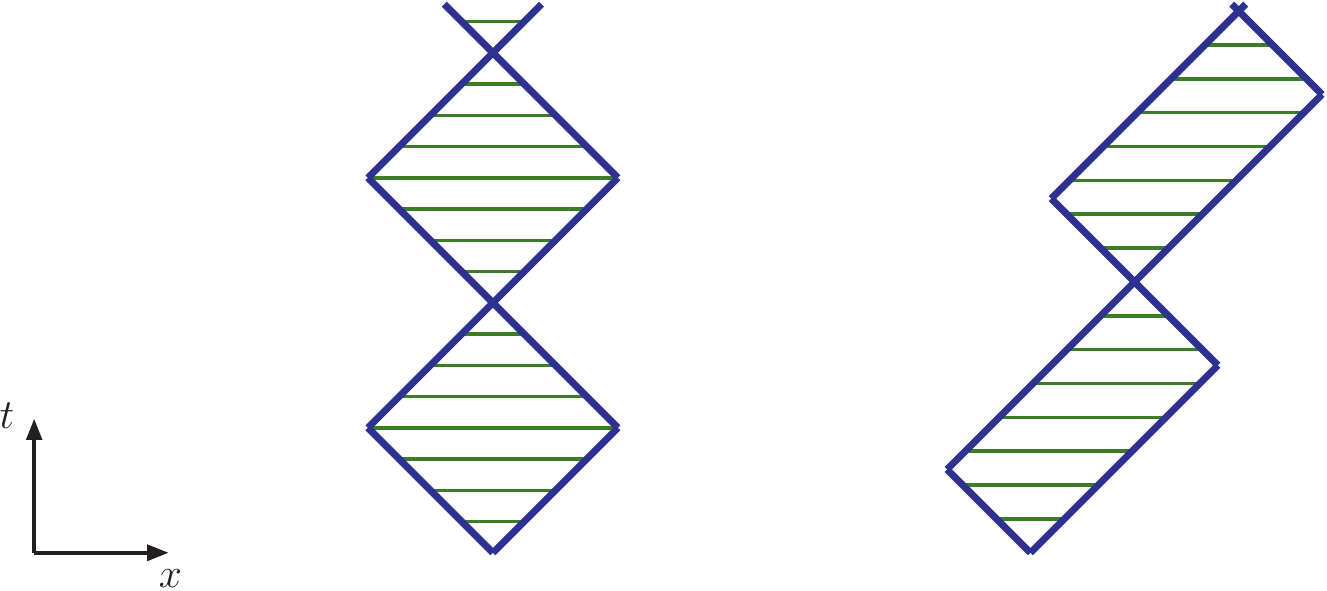}
\caption{Cartoon of string model, quark and antiquark moving apart from each other as seen in the meson rest frame (left) and in a boosted frame (right).}
\label{fig:string}
\end{figure}

The first step in understanding the structure of hadrons is to take this string picture very literally and to look at the space-time diagram of a meson ($q\bar{q}$), outlined in Figure~\ref{fig:string}. 

At some point in time the quark and the anti-quark (both assumed massless) are at the same point in space and they are flying apart at the speed of light, hence up a 45$^\circ$ diagonal in the space-time diagram. As they grow further apart they lay out a string between them that has a constant tension until the potential stored in this string uses up all their kinetic energy. At this point they turn around and the potential energy in the string accelerates the quarks towards each other until they meet back at the starting point, pass through each other, and the whole process starts over. These are the so-called ``yo-yo'' modes. 

It is a nice exercise in Lorentz transformations to think about what this process looks like when you are not in the rest frame but in a boosted frame, e.g. when the meson is moving to the right. The two points at which the quarks' directions reverse are not simultaneous and the string spends part of its cycle simply moving to the right, without expanding or contracting, and transferring momentum from one quark to the other. You can measure the speed of the meson from the slope of the line of its centre of mass. The area of these squares and rectangles is the same in both frames, i.e.\ it is Lorentz invariant, and obeys the area law
\begin{equation}
m^2=2\kappa^2 {\textrm{ area}}.\,
\end{equation}

The Lund String Model uses this picture as a model of hadronization. In the original, simple version of the model we start by ignoring gluon radiation. $e^+e^-$ annihilation is then a point-like source of $q\bar{q}$ pairs. In principle this system also has yo-yo modes, but in practice the space-time volume swept out is so large that another effect is able to dominate: in the intense chromomagnetic field of the string between the $q\bar{q}$ pair it is possible for additional $q\bar{q}$ pairs to spontaneously tunnel out of the vacuum. The chromomagnetic field is strong enough that the quarks are accelerated away from each other before they have time to re-annihilate\footnote{This is somewhat analogous to Hawking radiation.}. The effect is that the string separates into two strings: it breaks. By analogy with a similar process in QED you can estimate the probability of this happening, as
\begin{equation}
\frac{\d(\textrm{Probability})}{\d x\,\d t}\propto\exp{\frac{-\pi m^2_q}{\kappa}}\,.
\end{equation}
The mass dependence of this equation means that, for example, strange quarks will tunnel out less often than light quarks.

\begin{figure}
\centering
\includegraphics*[width=\textwidth,trim=0 50 0 0]{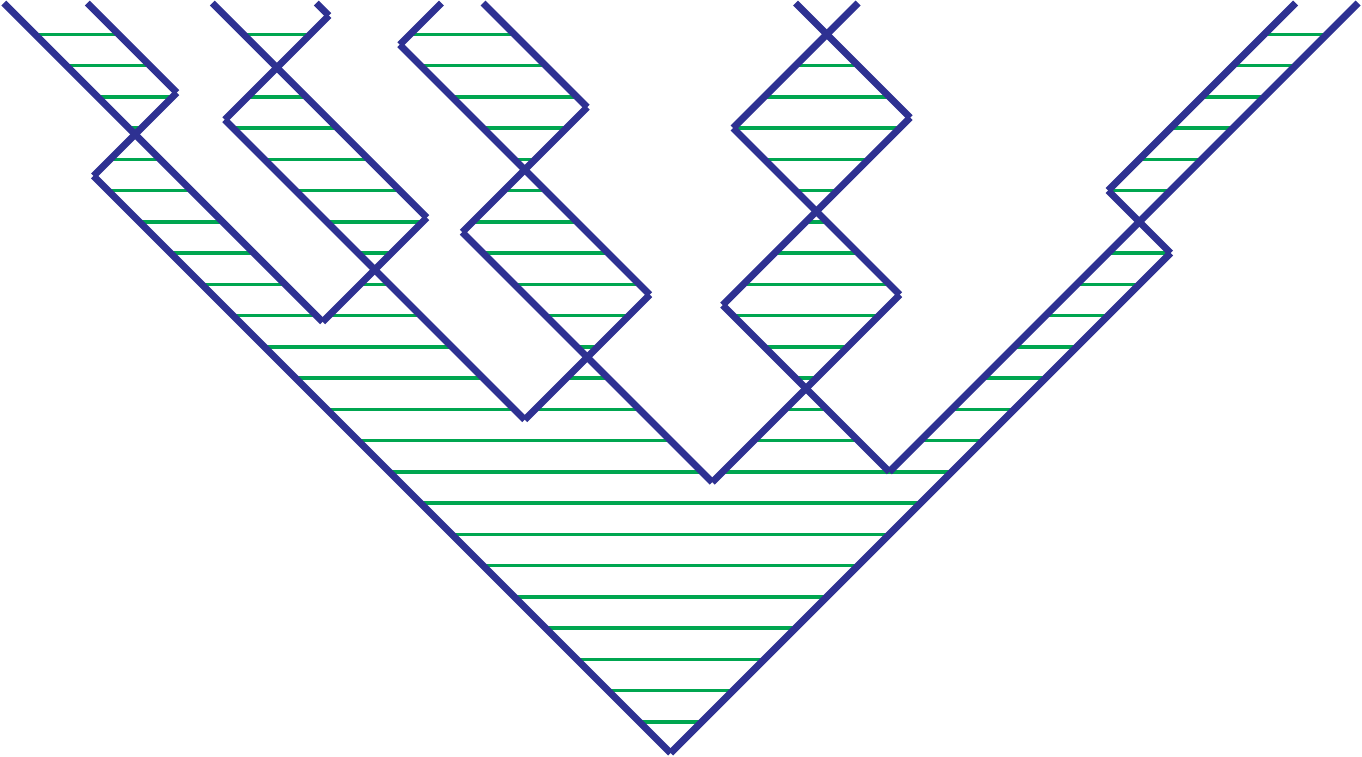}
\caption{Cartoon of the string corresponding to a high energy $q\bar{q}$ event breaking up into hadrons.}
\label{fig:fragmentation}
\end{figure}

When we have a lot of energy available it is likely that we will produce many hadrons, as the expanding string breaks into mesons long before the yo-yo point. Thus the original $q\bar{q}$ system has fragmented into a system of hadrons~-- this is the basic ingredient of the Lund String model illustrated in Figure~\ref{fig:fragmentation}.
The space time structure of this breakup is very interesting, as the breaks are causally disconnected, so they don't know about each other and there can be no causal correlations between them. The Lorentz invariance and acausality give strong constraints on this hadronization process. In the end, we get a fragmentation function for hadrons with a constrained form with two adjustable parameters, $a_\alpha$ and $a_\beta$,
\begin{equation}
f(z)\propto z^{a_\alpha-a_\beta-1}(1-z)^{a_\beta}\,.
\end{equation}
The tunnelling probability then becomes
\begin{equation}
\exp{-b(m^2_q+p^2_t)}\,,
\end{equation}
where the main tuneable parameters of the model are $a$, as described above, $b$, customarily called the ``Lund $b$ parameter'' related to the string tension, which can be seen to control the width of the $p_t$ distribution, and $m^2_q$, the masses of the individual quarks.

An important new feature of this model, relative to independent fragmentation, is that it is universal and predictive: having been tuned to describe quark events it can predict the confinement of gluons. This is again related to the colour structure. In a three parton ($q\bar{q}g$) system, the quark is colour-connected to the anticolour index of the gluon and the colour index of the gluon is connected to the antiquark. Thus the gluon makes a corner, or ``kink'', on the string. The acausality means that the breakup of the string is universal, but the Lorentz boost of a string means that the hadrons it produces go preferentially in the direction of its motion. Therefore most hadrons that the first string segment produces will go between the quark and the gluon, most hadrons from the second will go between the gluon and the antiquark and only very few hadrons will go between the quark and the antiquark. 

\begin{figure}
\centering
\hspace{\fill}
\includegraphics[width=.35\textwidth]{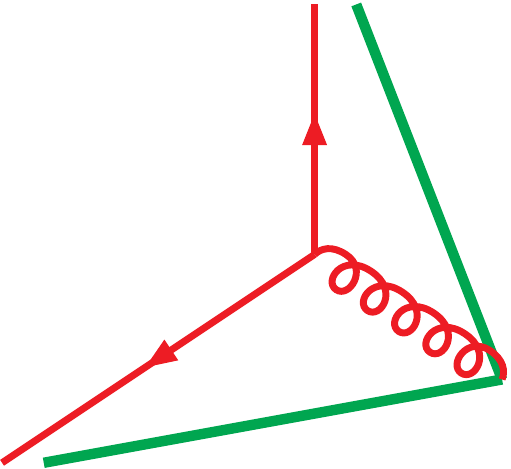}
\hspace{\fill}
\includegraphics[width=.35\textwidth]{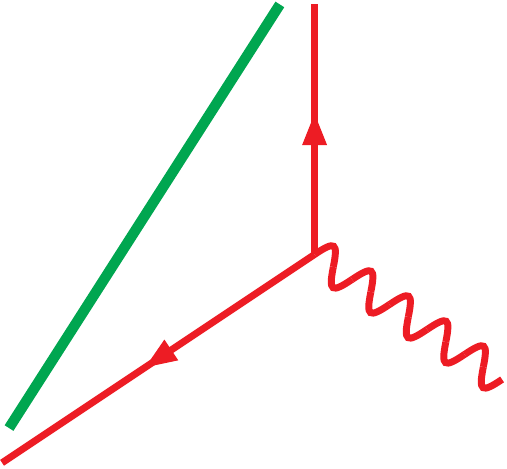}
\hspace{\fill}
\caption{String structure for $q\bar{q}g$ (left) and $q\bar{q}\gamma$ (right) events.}
\label{fig:string2}
\end{figure}

This definite prediction of the string model is known as the \emph{string effect} and can be seen experimentally, e.g.\ at the PETRA and LEP experiments, by comparing 3-jet events to 2-jet + photon events, which can be represented as in Figure~\ref{fig:string2}, where hadrons prefer to be between the quark and the antiquark.

In this model, there is a smooth matching with the parton shower, since a soft gluon with $k_\perp$ smaller than the inverse string width will have no effect on the hadronic final state.

In summary, the string model has a very strong physical motivation. It is a model of confinement, unlike the earlier independent fragmentation models. It is universal and gives the best description of data. However, for many of the effects for which it gives a strongly motivated qualitative prediction, in practice its quantitative prediction depends on free parameters that can be tuned to data. The smooth matching to the parton shower can also be seen as a disadvantage if one wishes to learn about the perturbative phase of QCD evolution as it, in a sense, can cover up the precise information from the parton shower. This motivated people to think of a new model which will be discussed in the following sections.

\subsection{Preconfinement and the Cluster Model}

In the planar, or large $N_c$, approximation, a gluon is a colour-anticolour pair. One can follow the colour structure of the parton shower and find for each external parton its colour partner to which it is colour connected. One finds that these colour-singlet pairs tend to end up close in phase space. The mass spectrum of colour-singlet pairs is asymptotically independent of energy or the production process and is peaked at low mass $\sim Q_0$. It depends on $Q_0$ and $\Lambda$, but not the shower scale $Q$. This property is known as preconfinement and is the inspiration for the cluster hadronization model.

The cluster model is motivated by thinking about the spectrum of mesonic states constructed from given quark and antiquark flavours. The lightest states are narrow, but the heavier ones are broad resonances~-- above 1.5~GeV or so one can picture a continuum of overlapping states of different spins. One can then think of the colour-anticolour pairs of preconfinement being projected directly onto this continuum. We call them clusters. These decay to lighter well-known resonances and stable hadrons. Once we have summed over all possible spins for a given process we effectively wash out all of the spin information and this assumption tells us that the decay should happen according to pure phase space. One immediate consequence of this is that heavier hadrons are suppressed~-- you get baryon and strangeness suppression ``for free'' (i.e.\ they are untuneable). The hadron-level properties are then fully determined by the cluster mass spectrum, i.e.\ by perturbative parameters. $Q_0$ is therefore a crucial parameter of the model.

This na\"{i}ve cluster model works well for the bulk of colour singlet states but, although the cluster mass spectrum is peaked at small mass, there is a broad tail to high masses. A small fraction of clusters are too heavy for isotropic two-body decay to be a reasonable approximation. In the cluster fission model, these high mass colour-anticolour pairs split into two lighter clusters in a longitudinal, i.e.\ rather string-like, way. The fission threshold becomes another crucial parameter for tuning the cluster model as, although only $\sim15\%$ of primary clusters get split, $\sim50\%$ of hadrons come from them.

The cluster model was found to describe data reasonably well, with far fewer parameters than the string model. However, although it was found to work well for the majority of hadrons, it was noticed that the leading hadrons were not hard enough. This was cured, at the expense of an additional parameter, by saying that perturbatively-produced quarks remember their direction somewhat, with probability distribution
\begin{equation}
P(\theta^2)\sim\exp{-\theta^2/\theta^2_0}\,,
\end{equation}
so that this cluster fragments more along its own axis. This again is not completely isotropic but more string-like as it remembers the direction along which the colour is expanding. It also has more adjustable parameters to fit the data.

The founding philosophy of the string and cluster model are quite opposite. The \emph{cluster model} emphasizes the perturbative phase of parton evolution and postulates that if this is correctly described, ``any old model'' of hadronization will be good enough. The \emph{string model} emphasizes the non-perturbative dynamics of the confinement of partons and started initially from a very simple treatment of the production of those partons. The accumulation of more precise data have led the models to converge to something more similar. The string model has had successively refined perturbative evolution and the cluster model has become successively more string-like. This leads one to wonder whether nature is pointing us towards a model in which the flavour mix is largely determined by the perturbative dynamics, as in the cluster model, and their distributions largely determined by non-perturbative string dynamics.

We close this section by commenting briefly on the universality of hadronization parameters. With so many free parameters, one might question the predictive power of these models. However, one finds in practice that the parameters are universal: that a single set of parameters describes the data at a wide range of energies and processes. One can show that this is a consequence of preconfinement: the perturbative production and evolution of partons takes care of the process- and energy-dependence and the transition from partons to hadrons is a local process, independent of these factors. Thus, hadronization models tuned to $e^+e^-$ annihilation, and lower energy hadron collider data, are highly predictive for LHC events.

\subsection{Secondary Decays}

An often underestimated ingredient of event generators is the model of secondary particle decays. This is more important than often realized because in both the string and the cluster models it is rare that the clusters decay directly to the stable pions and kaons seen in the detector. Mostly they decay to higher resonances which then decay further. One might say that these decays have been measured and one can just ``use the PDG'', but often not all resonances in a given multiplet have been measured, and rarely do the measured branching fractions add up to $100\%$ or respect expected symmetries such as isospin exactly. So when these data tables in the MC are built, a lot of choices need to be made. Moreover, in the case of multi-body decays, the matrix elements are highly non-trivial and appropriate models have to be constructed for them. The decay tables, and the decay models that implement them, actually have a significant effect on the hadron yields, transverse momentum release and hadronization correction to event shapes. The choice of decay table should therefore be considered as part of the tuned parameter set and a change of decay tables should be accompanied by a re-tune.

%% file: Section4.tex
The preceding steps of hard process, parton shower, hadronization and secondary decays are sufficient to fully describe the final state of the hard process, in which a high energy parton from each incoming hadron interact to produce an arbitrarily complex final state. However, this process involves the extraction of a coloured parton from each of the hadrons, which are colourless bound states of many coloured partons. We therefore have to consider how the hadron remnants evolve, hadronize and, potentially, interact with each other.

\begin{figure}
\centering
\includegraphics[width=1\textwidth]{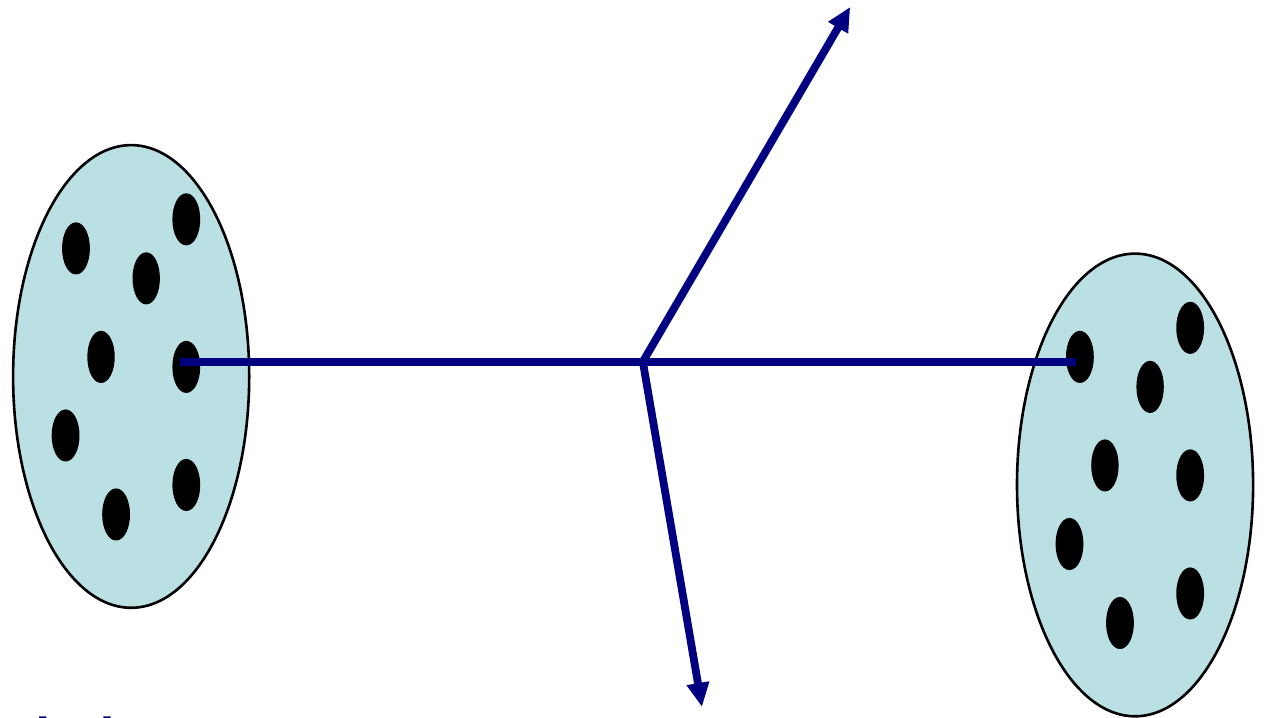}
\caption{Sketch of a proton-proton collision showing the Lorentz contraction of the protons.}
\label{fig:pancakes}
\end{figure}

In a proton's rest frame, it is a spherically symmetric extended object. Therefore in the laboratory frame where two protons collide at high energy, they look like extremely flattened discs due to Lorentz contraction, as shown in Figure~\ref{fig:pancakes}. Internal interactions are also extremely time dilated, so during the time that the discs overlap the protons' internal dynamics are effectively frozen. On the one hand, this means that a high energy interaction is extremely localized, and the whole of the parton shower and hadronization of the primary interaction happens in a very small space-time region and there is not time for it to be affected by the rest of the proton. On the other hand, it means that there is a very large overlap between the other partons in the protons and the possibility of additional interactions.

Historically, there are two main models that have been used. Even though the first is effectively ruled out by Tevatron and LHC data, it is still useful to discuss, to draw out the contrasting features of the second, more successful, model.

The \emph{non-perturbative model} is motivated by the fact that the soft parton-parton cross section is so large that there are many interactions everywhere in these discs and the assumption that these interactions are coherent across the discs. Thus the whole of one remnant interacts non-perturbatively with the whole of the other remnant. In the absence of an understanding of non-perturbative dynamics, our best hope is to parametrize data on these interactions. The only predictivity comes from the assumption that the underlying event at a given energy is independent of the hard process it underlies. This model was the default in \textsf{HERWIG} and is made available as an option in \textsf{Herwig++} as it is still interesting to have a ``straw man model'' of soft hadronic interactions without any hard component. However, all the models that successfully describe the LHC data have a perturbative origin.

In the \emph{perturbative models}, the idea is that the perturbative parton-parton cross section is so large that additional \emph{local} parton-parton interactions between other partons in the proton dominate. We do not therefore have a coherent scattering but multiple independent parton-parton interactions distributed across the disc, each producing their own hard processes and parton showers as well as the initial one that we started with.

The underlying event is closely linked with what are often called ``minimum bias'' events. These are the final states of a typical proton-proton collision and typically consist of a small number of hadrons at low transverse momentum distributed across a wide range of rapidities. Although the name ``minimum bias'' is widely used it is important to keep in mind that this is an experimental statement. By ``minimum'' we mean ``as little as possible'' so the amount of bias is dependent on the experiment. We would like to compare them with models that we think of as having zero bias, that are predicting all inelastic proton-proton collisions. To avoid confusion about the experiment-dependence of any minimum bias definition, the recent recommendation~\cite{guide} is to describe the event class as ``soft inclusive'' events, reserving the name ``minimum bias'' for experimental attempts to measure these events.

In analysis, people often assume that they can remove the effect of the underlying event by measuring soft inclusive events and then subtracting these off, as the features of the two are very similar. This works reasonably well as a first approximation but if you look into the details, fluctuations in the amount of underlying event and correlations between the underlying event and the measured jets are extremely important. Making this assumption can potentially be quite dangerous and it is possible to underestimate the size of underlying event corrections.

Most jet cross sections are very steep, typically falling like the $5^{th}$ or $6^{th}$ power of $p_t$. If jets get a little extra energy from the underlying event their distribution gets shifted sideways, but since the distribution is so steeply falling a small shift sideways corresponds to a very large upwards or downwards shift in the curve. So a small contamination from the underlying event can give a large change in the jet production rate with given kinematics. This means that jet cross sections are sensitive to rare fluctuations in the underlying event and just subtracting off an average amount of underlying event is not necessarily meaningful. Processes with different $p_t$ distributions will have different underlying event corrections~-- the steeper the $p_t$ distribution is, the more a jet sample will be populated by lower $p_t$ jets that have been shifted up by rare fluctuations in the underlying event. It is therefore extremely important to have reliable underlying event models that can predict this\footnote{The underlying event itself is not usually assumed to be correlated with the process but there is a trigger bias~-- if you look at jets in a given kinematic range the distribution of the primary jets determines how much they are affected by the underlying event. For example, in $Z$ and $ZZ$ production not much difference in underlying event is expected, but because the $ZZ$ process has a harder $p_t$ distribution of accompanying jets, it is less affected by the underlying event.}. The way to avoid this trouble is to not tune to the average amount of underlying event but to correct on an event-by-event basis, this way fluctuations and correlations will be better taken into account\cite{Cacciari:2007fd}.

\subsection{Multiparton Interaction Model}
 
\begin{figure}
\centering
\includegraphics[width=.7\textwidth]{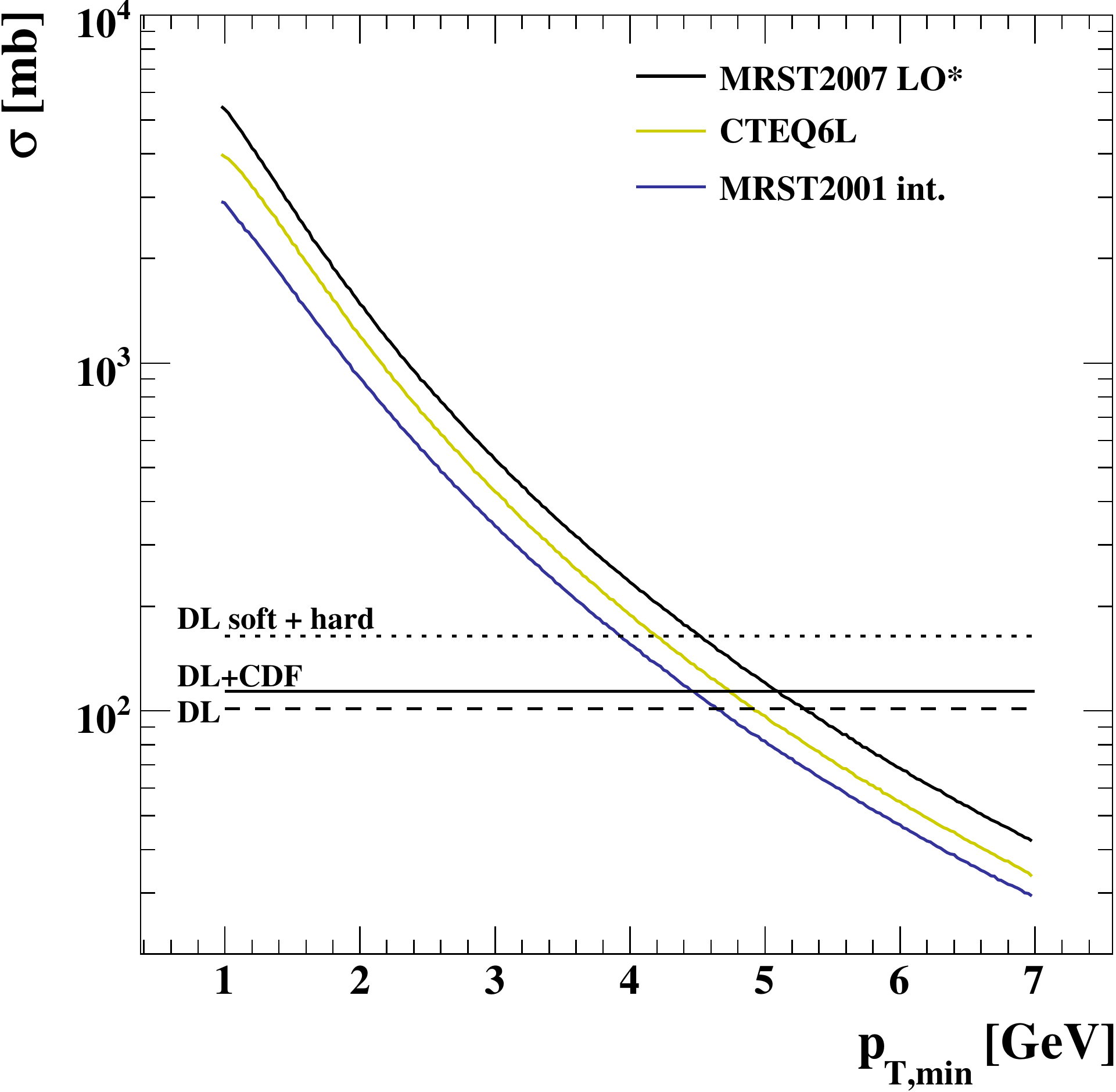}
\caption{Cross section at 14~TeV as a function of minimum $p_t$ for different PDF sets. Reproduced from~\cite{BBS}.}
\label{fig:MIModel}
\end{figure}

The starting point for the perturbative model is the observation that the hard parton-parton cross section is so large that we have many parton-parton collisions in one proton-proton collision. This is demonstrated in Figure~\ref{fig:MIModel}, which shows three curves using different PDF sets, with their different $\alpha_s$ values. The total proton-proton cross section predicted by three different models is shown for comparison\footnote{This figure was made for 14~TeV centre-of-mass energy and before the LHC measurements of the total cross section. The measured values at 7 and 8~TeV are closest to those predicted by the model labelled as ``DL+CDF'' in the figure.}.

For small $p_{t,min}$ and high energy the inclusive parton-parton cross section is larger than the total proton-proton cross section allowing more than one parton-parton scatter per proton-proton collision~\cite{MIM}. From PDFs calculated from deep inelastic scattering measurements, we know the distribution of momentum fractions of partons in the proton. What needs to be added is a model to describe the spatial distribution of partons within a proton. This is the only additional non-perturbative ingredient we need: with this we can calculate the distribution of number of scatters there are per proton-proton collision.

\begin{figure}
\centerline{%
\subfigure[]{\includegraphics[width=.55\textwidth]{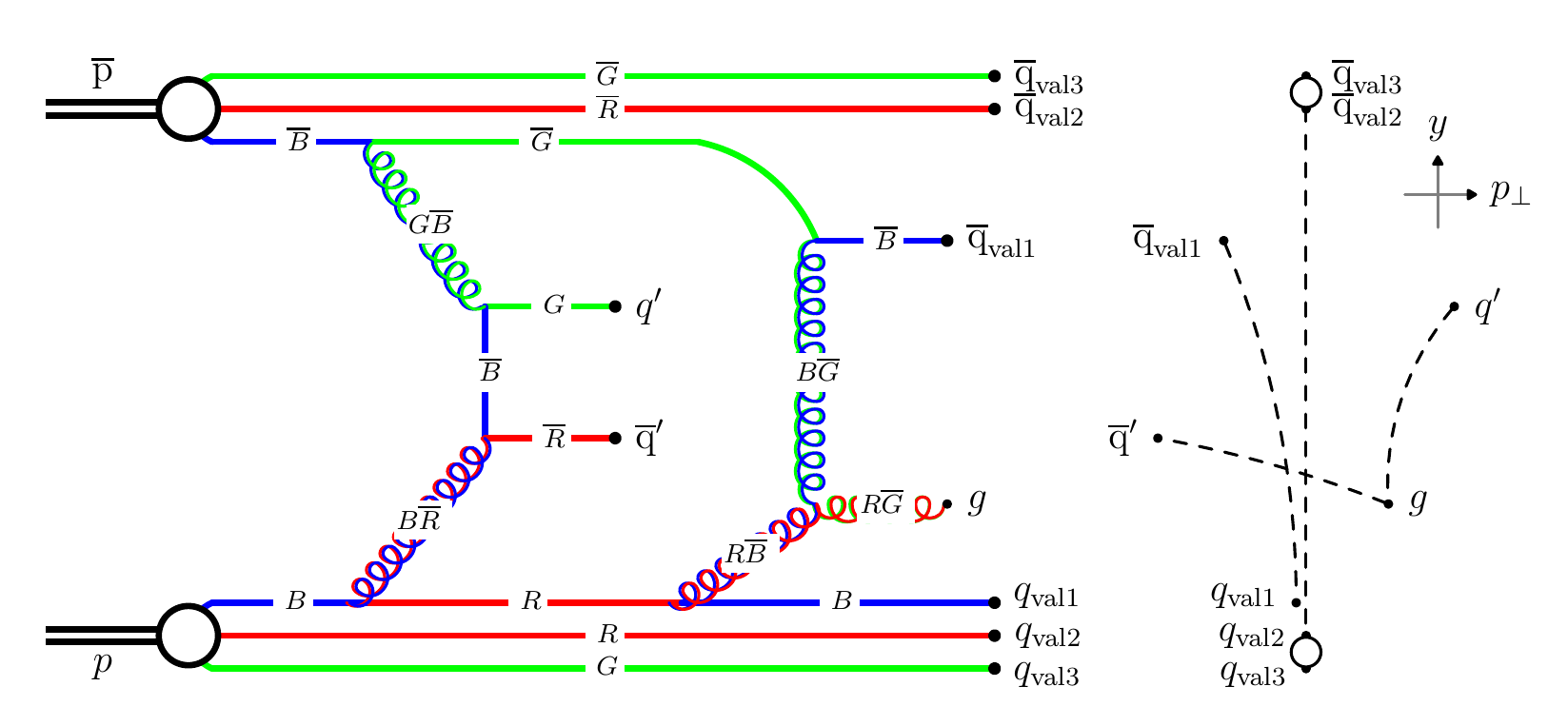}}
\subfigure[]{{\includegraphics[width=.55\textwidth]{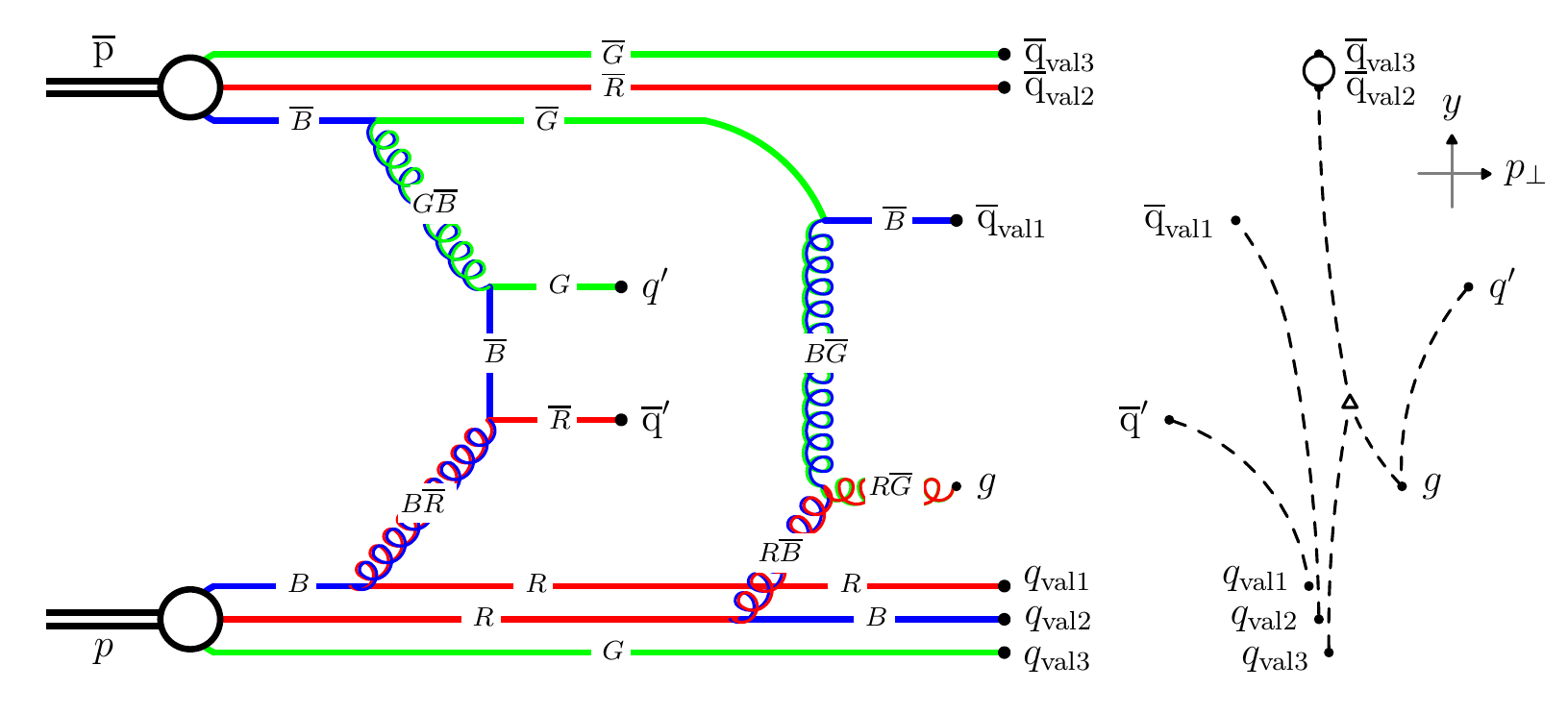}}}}
\caption{Example of colour correlations for two $p\bar{p}$ events with the same hard scatter colour structure. Reproduced from~\cite{guide}, adapted from \cite{Sjostrand:2004pf}.}
\label{fig:colorcorr}
\end{figure}

For these matter distributions, the current models usually make the assumption that $x$ and $b$ factorize:
\begin{equation}
\label{eq:xbfactorization}
n_i(x,b;\mu^2,s)=f_i(x;\mu^2)\,G(b,s)\,,
\end{equation}
with $f_i(x;\mu^2)$ the usual (inclusive) parton distribution functions, and that $n$-parton distributions are independent:
\begin{equation}
n_{i,j}(x_i,x_j,b_i,b_j)=n_i(x_i,b_i)\,n_j(x_j,b_j)\,,
\end{equation}
etc. In these approximations, the number of scatters at a fixed impact parameter, i.e.~with a given overlap between the two protons, is then given by a Poisson distribution. This can be integrated over impact parameter to calculate the $n$-scatter cross section:
\begin{equation}
\sigma_n=\int \d^2b\,\frac{\left(A(b)\sigma^{inc}\right)^n}{n!}\,\exp{\left(-A(b)\sigma^{inc}\right)}\,,
\end{equation}
with
\begin{equation}
A(b)=\int \d^2b_1\,G(b_1)\,\d^2b_2\,G(b_2)\,\delta^{(2)}(\mathbf{b}-\mathbf{b}_1+\mathbf{b}_2)\,.
\end{equation}
These ingredients are sufficient to generate a number of scatters and their kinematics and parton showers. The one remaining complication is how the colour of the different scatters is connected to each other. This has been studied by the \textsf{Pythia} authors in some detail\cite{Sjostrand:2004pf} and recently also by the \textsf{Herwig} authors\cite{Bahr:2008dy,creco}. Figure~\ref{fig:colorcorr} shows two $p\bar{p}$ events for which the hard processes ($gg\to q\bar{q}$ and $qg\to qg$) have exactly the same colour structure but the colour connections between the scatters and the external protons is different in the two cases. Although the parton showering will be identical in the two cases, the hadronization will differ, because the string connections (represented by the dashed lines in the right-hand part of each figure) differ.

Although perturbation theory can specify everything about the colour connections in the centre of the event, it doesn't tell us how these colours are hooked into the wavefunction of the original protons. So we need to supplement our model. The \textsf{Pythia} authors have studied different algorithms to do this in some detail and identified experimental observables that help to constrain them\cite{Skands:2007zg}.

\subsection{The Herwig++ Multiparton Interaction and Colour Reconnection Models}

The multiparton interaction model in \textsf{Herwig++} is developed from the one available as a plug-in to \textsf{HERWIG}, called \textsf{Jimmy}\cite{Butterworth:1996zw}, but with a number of new features\cite{Borozan:2002fk,Bahr:2008dy}. The idea is to use the eikonal model and optical theorem to connect the partonic scattering cross sections to the total, inelastic and elastic hadronic cross sections. A simple separation of partonic scatters is made into hard (above $p_{t,min}\sim3$--5~GeV), distributed perturbatively around `hot spots' of high parton density in the protons, and soft (below $p_{t,min}$), with a simple distribution with Gaussian\footnote{In fact, once the parameters are fixed, an interesting feature emerges: the width-squared of the Gaussian is forced to be negative, giving an ``inverted Gaussian'' with very few events at very low transverse momentum and a concentration of events around $p_{t,min}$. This lends support to the multiparton interaction model and the idea that the entire cross section could be described perturbatively at high energy.} transverse momentum and valence-like momentum fraction distributed across the whole of the protons' radii. Once the total cross section and elastic form factor are fixed by data there are only two free parameters: $p_{t,min}$ and the effective hot spot radius.

\begin{figure}
\centering
\includegraphics[width=1\textwidth]{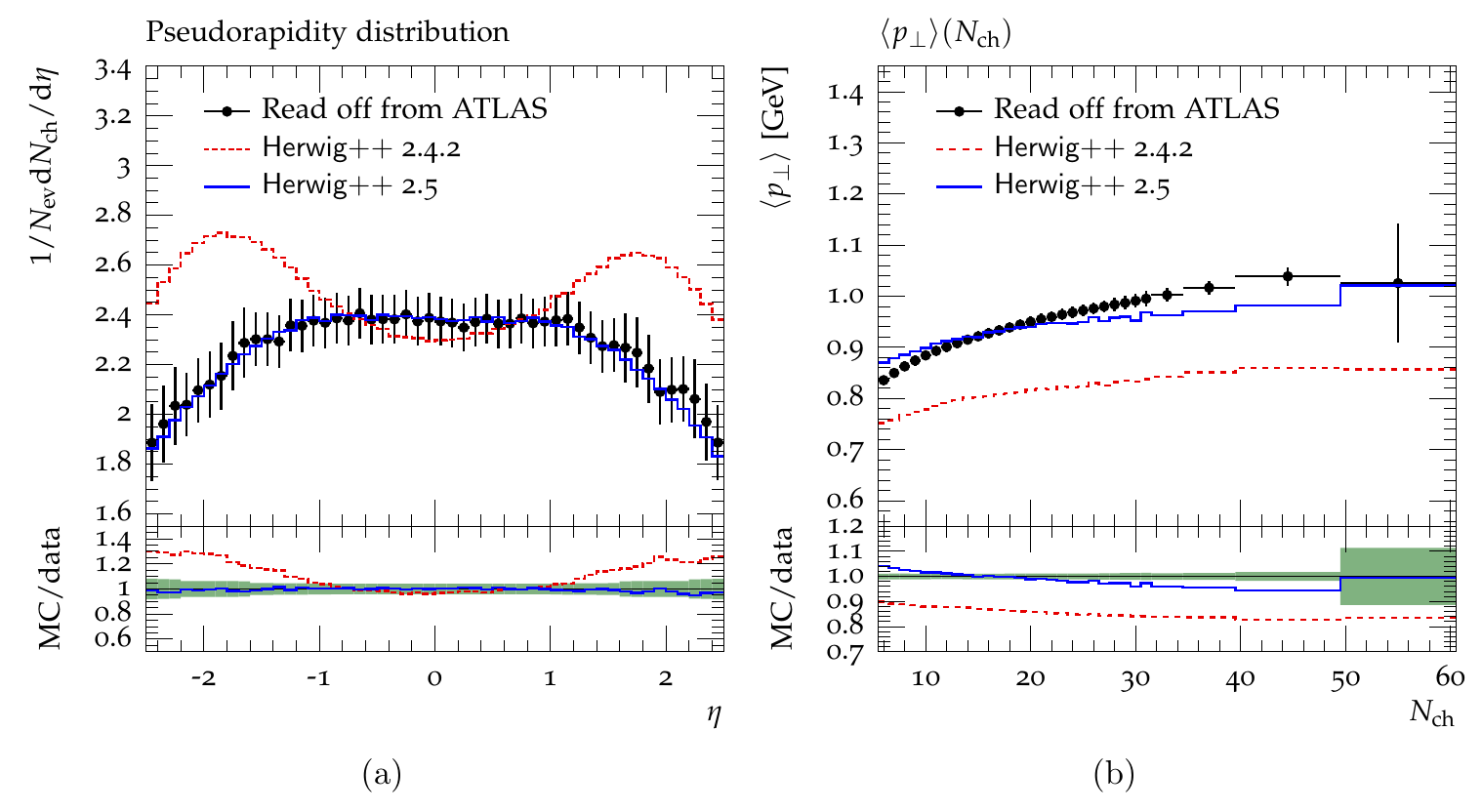}
\caption{Comparison of \textsf{Herwig++ 2.4.2}, without colour reconnections, and \textsf{Herwig++ 2.5}, with colour reconnections, to ATLAS minimum-bias distributions at $\sqrt{s} = 0.9$ TeV with $N_{ch}\ge6$, \mbox{$p_t > 500$ MeV} and $|\eta| < 2.5$, reproduced from~\cite{creco}.}
\label{fig:colorreco}
\end{figure}

When first implemented, this model gave a good description of underlying event data, but failed badly for soft inclusive analyses. It was realized that this was due to the issue of colour correlations between the scatters, which was not very carefully handled in the first implementation. R\"{o}hr, Siodmok and Gieseke have implemented a new model of colour reconnections in \textsf{Herwig++}\cite{creco} based on the momentum structure. This also gives reconnection effects in $e^+e^-$ annihilation, so a refit of LEP-I and LEP-II data is necessitated, but the conclusion is that one can get a good fit of the $e^+e^-$ and LHC data, see for example Figure~\ref{fig:colorreco} where the red line is the best fit to the data without colour reconnections and is clearly nothing like the data. An important conclusion of this study is that the hadronization parameters are correlated with the reconnection probability~-- changing one necessitates a retuning of the other, but a good fit can be obtained for a wide range of colour reconnection probabilities.

\subsection{Pythia implementation}

\textsf{Pythia} was the first event generator to incorporate multiparton interactions and its implementation is very well developed. An interesting feature that has emerged, that is not shared by \textsf{Herwig} or \textsf{Sherpa}, is the possibility that through colour connection effects there can be an interplay between multiparton interactions and the initial-state parton shower\cite{Sjostrand:2004ef}. Starting with a single hard interaction at some value of $p_t$, the simulation evolves downwards in $p_t$, with the possibility at every step of either generating an emission from an incoming parton or an additional scatter. At any point during this evolution there can be colour cross-talk between these different interactions and this will affect the distribution of hadronization and of the partons that are radiated by these multiple scatters.

A recent study\cite{Corke:2009tk} also considered rescattering, where two partons out of one proton can scatter with the same parton from the other. This is suppressed by the fact that it must be a local process: the two partons in one proton must both be overlapping with the other. Nevertheless it does give another contribution to the fluctuations in the underlying event.

Most existing models make the simplifying assumption in Equation (\ref{eq:xbfactorization}) of factorization of $x$ and $b$, i.e.\ of the momentum and spatial distribution of partons. In another recent study\cite{Corke:2011yy}, Corke and Sj\"ostrand implemented a model without this assumption by considering a Gaussian matter distribution with an $x$-dependent width,
\begin{equation}
a(x)=a_0\left(1+a_1\ln\frac{1}{x}\right),
\end{equation}
and looked at what you could learn about these parameters from the data. The effect is to start producing more correlation with the underlying event. A higher mass final state is produced at higher $x$, therefore it has a narrower matter distribution and more underlying event. There is a correlation between the momentum used in the hard collision and the underlying event that accompanies it. They compared the underlying event in $Z$ events and in events that produce a $Z'$ of 1 TeV and found significant differences. They could obtain equally good fits of the existing underlying event data but with significant differences in their extrapolation to higher-scale processes. This is clearly something that requires further study, to improve the models and to understand the uncertainty they introduce in high-mass searches using jets, for example.

\subsection{Underlying Event Measurements}

Despite a $\sim$ 25 year history, many aspects of our understanding of multi-parton interactions are still in their infancy. The Tevatron and especially LHC experiments have already opened up huge areas for further study, not only with a big step up in energy but also with much higher efficiency, purity and phase space coverage than the previous measurements. There has also been a big change in the culture around the measurements, with an emphasis on physical (experiment-independent and generator-independent) observables that can be directly compared between experiments and with a wide variety of models, now or in the future. There is also a move towards making more targeted measurements of observables that are sensitive to specific physical effects, such as colour reconnections. The general conclusion is that all the existing models can describe the general underlying event and soft inclusive data well with tuning. The emphasis is moving towards understanding of correlations between hard and underlying events, rare corners of phase space (such as high multiplicity soft events) and the relationships between different model components. One of the main motivations for these studies is the fact that jet corrections depend strongly on these correlations, and high moments of distributions, and are physics-process dependent. A deeper understanding, and greater predictivity, is still needed.

\section{Summary}
\label{sec:5}

As a summary of our discussion of event generators, we recall the main subjects that we have covered, commenting on how well they are understood from first principles. We briefly touched on the hard process which is generally a direct implementation of tree-level perturbation theory and hence extremely well understood. We discussed in detail the parton shower which is an approximation to all-order perturbation theory and therefore in principle well understood. Various approximations are made in constructing parton showers and the effect of these is not always as small as anticipated. The cutting edge here is the matching between higher order fixed-order perturbation theory and parton showers, which should, in principle, be fully understandable from perturbation theory, but is at present the subject of some uncertainty. We then talked about hadronization which is less well understood from first principles. Although there are different models, they are well constrained by data and the extrapolation to LHC data is considered to be fairly reliable. Lastly, the underlying event is the least well understood out of all these. It is only weakly constrained by previous data and different models that fit the available data give quite different extrapolations. Moreover, it is important to recall that correlations and rare fluctuations in the underlying event are as important as its average properties and are even less well tied down.

Monte Carlo event generators are increasingly used as tools in almost every aspect of high energy collider physics. As the data become more precise it becomes increasingly important not to use them as black boxes, but to question how reliable they are for the application at hand. The important question to ask is ``What physics is dominating my effect?''. We hope that these lecture notes have helped equip the reader to answer this question.

%% file: referenc.tex
%
%
%

%% file: Seymour.bbl
\begin{thebibliography}{99.}%
%
%
%
%

\bibitem{Aad:2012tfa}
G.~Aad {\it et al.}  [ATLAS Collaboration],
  ``Observation of a new particle in the search for the Standard Model Higgs boson with the ATLAS detector at the LHC'',
  Phys.\ Lett.\ B {\bf 716} (2012) 1
  [arXiv:1207.7214 [hep-ex]].

\bibitem{Chatrchyan:2012ufa}
S.~Chatrchyan {\it et al.}  [CMS Collaboration],
  ``Observation of a new boson at a mass of 125 GeV with the CMS experiment at the LHC'',
  Phys.\ Lett.\ B {\bf 716} (2012) 30
  [arXiv:1207.7235 [hep-ex]].

\bibitem{guide}
A.~Buckley, J.~Butterworth, S.~Gieseke, D.~Grellscheid, S.~H{\"o}che, H.~Hoeth, F.~Krauss and L.~L{\"o}nnblad {\it et al.},
  ``General-purpose event generators for LHC physics'',
  Phys.\ Rept.\  {\bf 504} (2011) 145
  [arXiv:1101.2599 [hep-ph]].

\bibitem{Ellis:1991qj}
  R.~K.~Ellis, W.~J.~Stirling and B.~R.~Webber,
  ``QCD and collider physics'',
  Camb.\ Monogr.\ Part.\ Phys.\ Nucl.\ Phys.\ Cosmol.\  {\bf 8} (1996) 1.

\bibitem{Abe:1994nj}
F.~Abe {\it et al.}  [CDF Collaboration],
  ``Evidence for color coherence in $p\bar{p}$ collisions at $\sqrt{s} = 1.8$ TeV'',
  Phys.\ Rev.\ D {\bf 50} (1994) 5562.

\bibitem{Catani:2002hc}
  S.~Catani, S.~Dittmaier, M.~H.~Seymour and Z.~Trocsanyi,
  ``The Dipole formalism for next-to-leading order QCD calculations with massive partons'',
  Nucl.\ Phys.\ B {\bf 627} (2002) 189
  [hep-ph/0201036].

\bibitem{Norrbin:2000uu}
  E.~Norrbin and T.~Sj{\"o}strand,
  ``QCD radiation off heavy particles'',
  Nucl.\ Phys.\ B {\bf 603} (2001) 297
  [hep-ph/0010012].

\bibitem{NS}
Z.~Nagy and D.~E.~Soper,
  ``Matching parton showers to NLO computations'',
  JHEP {\bf 0510} (2005) 024
  [hep-ph/0503053].

\bibitem{DTW}
M.~Dinsdale, M.~Ternick and S.~Weinzierl,
  ``Parton showers from the dipole formalism'',
  Phys.\ Rev.\ D {\bf 76} (2007) 094003
  [arXiv:0709.1026 [hep-ph]].

\bibitem{Sherpa1}
S.~Schumann and F.~Krauss,
  ``A Parton shower algorithm based on Catani-Seymour dipole factorisation'',
  JHEP {\bf 0803} (2008) 038
  [arXiv:0709.1027 [hep-ph]].

\bibitem{Sherpa2}
J.-C.~Winter and F.~Krauss,
  ``Initial-state showering based on colour dipoles connected to incoming parton lines'',
  JHEP {\bf 0807} (2008) 040
  [arXiv:0712.3913 [hep-ph]].

\bibitem{Vincia}
W.~T.~Giele, D.~A.~Kosower and P.~Z.~Skands,
  ``A Simple shower and matching algorithm'',
  Phys.\ Rev.\ D {\bf 78} (2008) 014026
  [arXiv:0707.3652 [hep-ph]].

\bibitem{CS}
  S.~Catani and M.~H.~Seymour,
  ``The Dipole formalism for the calculation of QCD jet cross-sections at next-to-leading order'',
  Phys.\ Lett.\ B {\bf 378} (1996) 287
  [hep-ph/9602277].
\\
  S.~Catani and M.~H.~Seymour,
  ``A General algorithm for calculating jet cross-sections in NLO QCD'',
  Nucl.\ Phys.\ B {\bf 485} (1997) 291
   [Erratum-ibid.\ B {\bf 510} (1998) 503]
  [hep-ph/9605323].

\bibitem{DK}
D.~A.~Kosower,
  ``Antenna factorization of gauge theory amplitudes'',
  Phys.\ Rev.\ D {\bf 57} (1998) 5410
  [hep-ph/9710213].

\bibitem{Catani:2001cc}
  S.~Catani, F.~Krauss, R.~Kuhn and B.~R.~Webber,
  ``QCD matrix elements + parton showers'',
  JHEP {\bf 0111} (2001) 063
  [hep-ph/0109231].

\bibitem{Krauss:2002up}
  F.~Krauss,
  ``Matrix elements and parton showers in hadronic interactions'',
  JHEP {\bf 0208} (2002) 015
  [hep-ph/0205283].

\bibitem{Lonnblad:2001iq}
  L.~L\"onnblad,
  ``Correcting the color dipole cascade model with fixed order matrix elements'',
  JHEP {\bf 0205} (2002) 046
  [hep-ph/0112284].

\bibitem{Alwall:2007fs}
  J.~Alwall, S.~H\"oche, F.~Krauss, N.~Lavesson, L.~L\"onnblad, F.~Maltoni, M.~L.~Mangano and M.~Moretti {\it et al.},
  ``Comparative study of various algorithms for the merging of parton showers and matrix elements in hadronic collisions'',
  Eur.\ Phys.\ J.\ C {\bf 53} (2008) 473
  [arXiv:0706.2569 [hep-ph]].

\bibitem{Frixione:2002ik}
  S.~Frixione and B.~R.~Webber,
  ``Matching NLO QCD computations and parton shower simulations'',
  JHEP {\bf 0206} (2002) 029
  [hep-ph/0204244].

\bibitem{Nason:2006hfa}
  P.~Nason and G.~Ridolfi,
  ``A Positive-weight next-to-leading-order Monte Carlo for Z pair hadroproduction'',
  JHEP {\bf 0608} (2006) 077
  [hep-ph/0606275].

\bibitem{Lavesson:2008ah}
  N.~Lavesson and L.~L\"onnblad,
  ``Extending CKKW-merging to One-Loop Matrix Elements'',
  JHEP {\bf 0812} (2008) 070
  [arXiv:0811.2912 [hep-ph]].

\bibitem{Hamilton:2010wh}
  K.~Hamilton and P.~Nason,
  ``Improving NLO-parton shower matched simulations with higher order matrix elements'',
  JHEP {\bf 1006} (2010) 039
  [arXiv:1004.1764 [hep-ph]].

\bibitem{Hoche:2010kg}
  S.~H\"oche, F.~Krauss, M.~Sch\"onherr and F.~Siegert,
  ``NLO matrix elements and truncated showers'',
  JHEP {\bf 1108} (2011) 123
  [arXiv:1009.1127 [hep-ph]].

\bibitem{Sjostrand:2000wi}
  T.~Sj\"ostrand, P.~Eden, C.~Friberg, L.~L\"onnblad, G.~Miu, S.~Mrenna and E.~Norrbin,
  ``High-energy physics event generation with PYTHIA 6.1'',
  Comput.\ Phys.\ Commun.\  {\bf 135} (2001) 238
  [hep-ph/0010017].

\bibitem{Corcella:2000bw}
  G.~Corcella, I.~G.~Knowles, G.~Marchesini, S.~Moretti, K.~Odagiri, P.~Richardson, M.~H.~Seymour and B.~R.~Webber,
  ``HERWIG 6: An Event generator for hadron emission reactions with interfering gluons (including supersymmetric processes)'',
  JHEP {\bf 0101} (2001) 010
  [hep-ph/0011363].

\bibitem{Lonnblad:1992tz}
  L.~L\"onnblad,
  ``ARIADNE version 4: A Program for simulation of QCD cascades implementing the color dipole model'',
  Comput.\ Phys.\ Commun.\  {\bf 71} (1992) 15.

\bibitem{Sjostrand:2006za}
  T.~Sj\"ostrand, S.~Mrenna and P.~Z.~Skands,
  ``PYTHIA 6.4 Physics and Manual'',
  JHEP {\bf 0605} (2006) 026
  [hep-ph/0603175].

\bibitem{Sjostrand:2007gs}
  T.~Sj\"ostrand, S.~Mrenna and P.~Z.~Skands,
  ``A Brief Introduction to PYTHIA 8.1'',
  Comput.\ Phys.\ Commun.\  {\bf 178} (2008) 852
  [arXiv:0710.3820 [hep-ph]].

\bibitem{Gleisberg:2008ta}
  T.~Gleisberg, S.~.H\"oche, F.~Krauss, M.~Sch\"onherr, S.~Schumann, F.~Siegert and J.~Winter,
  ``Event generation with SHERPA 1.1'',
  JHEP {\bf 0902} (2009) 007
  [arXiv:0811.4622 [hep-ph]].

\bibitem{Bahr:2008pv}
  M.~B\"ahr, S.~Gieseke, M.~A.~Gigg, D.~Grellscheid, K.~Hamilton, O.~Latunde-Dada, S.~Pl\"atzer and P.~Richardson {\it et al.},
  ``Herwig++ Physics and Manual'',
  Eur.\ Phys.\ J.\ C {\bf 58} (2008) 639
  [arXiv:0803.0883 [hep-ph]].

\bibitem{Platzer:2011bc}
  S.~Pl\"atzer and S.~Gieseke,
  ``Dipole Showers and Automated NLO Matching in Herwig++'',
  Eur.\ Phys.\ J.\ C {\bf 72} (2012) 2187
  [arXiv:1109.6256 [hep-ph]].

\bibitem{Cacciari:2007fd}
  M.~Cacciari and G.~P.~Salam,
  ``Pileup subtraction using jet areas'',
  Phys.\ Lett.\ B {\bf 659} (2008) 119
  [arXiv:0707.1378 [hep-ph]].

\bibitem{BBS}
M. B\"{a}hr, J.~M.~Butterworth and M.~H.~Seymour,
  ``The Underlying Event and the Total Cross Section from Tevatron to the LHC'',
  JHEP {\bf 0901} (2009) 065
  [arXiv:0806.2949 [hep-ph]].

\bibitem{MIM}
T. Sj\"{o}strand and M.~van Zijl,
  ``A Multiple Interaction Model for the Event Structure in Hadron Collisions'',
  Phys.\ Rev.\ D {\bf 36} (1987) 2019.

\bibitem{Sjostrand:2004pf}
  T.~Sj\"ostrand and P.~Z.~Skands,
  ``Multiple interactions and the structure of beam remnants'',
  JHEP {\bf 0403} (2004) 053
  [hep-ph/0402078].

\bibitem{Bahr:2008dy}
  M.~B\"ahr, S.~Gieseke and M.~H.~Seymour,
  ``Simulation of multiple partonic interactions in Herwig++'',
  JHEP {\bf 0807} (2008) 076
  [arXiv:0803.3633 [hep-ph]].

\bibitem{creco}
S. Gieseke, C.A. R\"{o}hr, A. Si\'{o}dmok,
  ``Colour reconnections in Herwig++'',
  Eur.\ Phys.\ J.\ C {\bf 72} (2012) 2225
  [arXiv:1206.0041 [hep-ph]].

\bibitem{Skands:2007zg}
  P.~Z.~Skands and D.~Wicke,
  ``Non-perturbative QCD effects and the top mass at the Tevatron'',
  Eur.\ Phys.\ J.\ C {\bf 52} (2007) 133
  [hep-ph/0703081 [HEP-PH]].

\bibitem{Butterworth:1996zw}
  J.~M.~Butterworth, J.~R.~Forshaw and M.~H.~Seymour,
  ``Multiparton interactions in photoproduction at HERA'',
  Z.\ Phys.\ C {\bf 72} (1996) 637
  [hep-ph/9601371].

\bibitem{Borozan:2002fk}
  I.~Borozan and M.~H.~Seymour,
  ``An Eikonal model for multiparticle production in hadron hadron interactions'',
  JHEP {\bf 0209} (2002) 015
  [hep-ph/0207283].

\bibitem{Sjostrand:2004ef}
  T.~Sj{\"o}strand and P.~Z.~Skands,
  ``Transverse-momentum-ordered showers and interleaved multiple interactions'',
  Eur.\ Phys.\ J.\ C {\bf 39} (2005) 129
  [hep-ph/0408302].

\bibitem{Corke:2009tk}
  R.~Corke and T.~Sj{\"o}strand,
  ``Multiparton Interactions and Rescattering'',
  JHEP {\bf 1001} (2010) 035
  [arXiv:0911.1909 [hep-ph]].

\bibitem{Corke:2011yy}
R. Corke and T. Sj\"{o}strand,
  ``Multiparton Interactions with an x-dependent Proton Size'',
  JHEP {\bf 1105} (2011) 009
  [arXiv:1101.5953 [hep-ph]].

\end{thebibliography}
